\documentclass[prepreint,twocolumn]{aastex63}
\hypersetup{linkcolor={blue},citecolor={blue}}
\usepackage{cprotect}

\received{\today}
\revised{\today}
\accepted{\today}
\submitjournal{ApJ}

\shorttitle{GRB cocoon}
\shortauthors{Suzuki \& Maeda}

\begin{document}

\title{Chemical stratification in a long gamma-ray burst cocoon and early-time spectral signatures of supernovae associated with gamma-ray bursts}

\correspondingauthor{Akihiro Suzuki}
\email{akihiro.suzuki@nao.ac.jp}

\author[0000-0002-7043-6112]{Akihiro Suzuki}
\affiliation{Division of Science, National Astronomical Observatory of Japan, 2-21-1 Osawa, Mitaka, Tokyo 181-8588, Japan}



\author[0000-0003-2611-7269]{Keiichi Maeda}
\affil{Department of Astronomy, Kyoto University, Kitashirakawa-Oiwake-cho, Sakyo-ku, Kyoto 606-8502, Japan}








\begin{abstract}
We present the results of 3D hydrodynamic simulations of gamma-ray burst (GRB) jet emanating from a massive star with a particular focus on the formation of high-velocity quasi-spherical ejecta and the jet-induced chemical mixing. 
Recent early-time optical observations of supernovae associated with GRBs (e.g., GRB 171205A/SN 2017iuk) indicate a considerable amount of heavy metals in the high-velocity outer layers of the ejecta. 
Using our jet simulations, we show that the density and chemical structure of the outer ejecta implied by observations can be naturally reproduced by a powerful jet penetrating the progenitor star. 
We consider three representative jet models with a stripped massive star, a standard jet, a weak jet, and a jet choked by an extended circumstellar medium, to clarify the differences in the dynamical evolution and the chemical properties of the ejected materials. 
The standard jet successfully penetrates the progenitor star and creates a quasi-spherical ejecta component (cocoon). 
The jet-induced mixing significantly contaminates the cocoon with heavy elements that have been otherwise embedded in the inner layer of the ejecta.  
The weak and choked jet models fail to produce an ultra-relativistic jet but produce a quasi-spherical cocoon with different chemical properties. 
We discuss the impact of the different jet-star interactions on the expected early-time electromagnetic signatures of long GRBs and how to probe the jet dynamics from observations. 
\end{abstract}



\section{Introduction}\label{sec:introduction} 
The terminal explosions of massive stars, i.e., core-collapse supernovae (SNe), are occasionally accompanied by ultra-relativistic jets outshining across the electromagnetic spectrum. 
The gamma-ray emission from such a powerful jet is observed as a burst-like gamma-ray point source, which is called a gamma-ray burst (GRB) and is extensively studied since its discovery \citep{1999PhR...314..575P,2004RvMP...76.1143P,2006RPPh...69.2259M,2015PhR...561....1K}. 
One of the important progress in this field has been made when the association of highly energetic SNe with long-duration GRBs was revealed \citep[e.g.,][]{2006ARA&A..44..507W,2013RSPTA.37120275H,2017AdAst2017E...5C}. 
The firmly established GRB-SN connection suggests a close physical link between the massive stellar collapse and the formation of a relativistic jet. 
There have been, however, many unsolved problems, such as the progenitor system responsible for long GRBs, the central compact object, and the jet driving mechanism.

Observations of SNe associated with GRBs (GRB-SNe) have provided vital information on these issues.  
From the observational viewpoint, GRB-SNe are exclusively classified into a special sub-class of Type-Ic SNe with the absence of hydrogen and helium spectral lines and the presence of broad absorption features (the so-called broad-lined Type-Ic SNe or SNe Ic-BL for short). 
The light curve modelings and spectral analyses of GRB-SNe suggest a bare carbon-oxygen (CO) core exploding with an energy $\sim10$ times higher (i.e., $\sim 10^{52}$ erg) than the canonical SN explosion energy. 
This picture implies the most massive stars having lost their hydrogen and helium envelopes as the GRB progenitor. 
The highly energetic nature of GRB-SNe also suggests that the explosion mechanism should be different from the canonical neutrino-aided explosion for normal CCSNe \citep[e.g.,][]{2012ARNPS..62..407J} and is somehow related with the jet formation.  
This poses one of the biggest theoretical challenges in GRBs and associated SNe; how to explain the extreme explosion energy. 
The GRB central engine has been claimed to be either a black hole-accretion disk system \citep[the so-called collapsar scenario;][]{1999ApJ...524..262M} or a rapidly rotating proto-magnetar \citep{1992Natur.357..472U,2011MNRAS.413.2031M}. 
However, there has been no consensus on how exactly the central compact object can power the ultra-relativistic jet and the bright SN. 

Spectral analyses of GRB-SNe have provided rich information on their origin \citep[e.g.,][]{2000ApJ...545..407M,2001ApJ...550..991N,2016ApJ...832..108M}. 
In particular, the chemical tomography has been possible for nearby GRB-SNe, such as GRB 980424/SN 1998bw \citep{1998Natur.395..663K,1998Natur.395..670G}, GRB 060218/SN 2006aj \citep{2006Natur.442.1008C,2006Natur.442.1018M,2006ApJ...645L..21M,2006Natur.442.1014S}, and GRB 100316D/SN 2010bh \citep{2011MNRAS.411.2792S,2011ApJ...740...41C,2012ApJ...753...67B,2012A&A...539A..76O}. 
These nearby GRB-SNe are usually accompanied by low-luminosity GRBs and admittedly their origin may be different from more energetic GRBs found at cosmological distances. 
Nevertheless, scrutinizing these nearby GRB-SNe has helped us develop the physical picture of energetic cosmic explosions from the most massive stars.

One of the recent progress in nearby GRB-SNe studies is the realization of photometric and spectroscopic follow-up observations as early as a day after gamma-ray detection. 
An SN ejecta gradually reveals itself from outer to inner layers as it becomes transparent. 
Early-time observations of GRB-SNe therefore unveil the spectroscopic appearance of the outermost layer of the ejecta above the photosphere. 
For example, the follow-up optical spectroscopic observations of a nearby GRB 161219B/SN 2016jca at a redshift $z=0.1475$ were conducted about 2 days after the gamma-ray detection \citep{2017A&A...605A.107C,2019MNRAS.487.5824A} along with late-time observations in other wavelengths \citep[e.g.,][]{2018ApJ...862...94L,2019ApJ...870...67A}. 
\cite{2019MNRAS.487.5824A} conducted spectral analyses of the optical spectra and found a high Ni abundance in the high-velocity outermost layer ($\simeq 35,000\,\mathrm{km}\,\mathrm{s}^{-1}$), which decreases toward inner slower layers ($\simeq 10,000\,\mathrm{km}\,\mathrm{s}^{-1}$). 
They suggest that the early optical light is emitted from a photosphere located in a wide jet with a large fraction of heavy elements. 

More striking evidence of heavy elements in early-time GRB-SN spectra was obtained in the follow-up observations of GRB 171205A/SN 2017iuk \citep{2018A&A...619A..66D,2018ApJ...867..147W,2019Natur.565..324I}. 
The optical observations started $< 1$ day after the gamma-ray detection \citep{2019Natur.565..324I} identified high-velocity ejecta with an expansion velocity higher than $0.3c$ and blue-shifted absorption features of explosive nucleosynthesis products, such as Ca and Si. 
Their detailed spectral modelings revealed the presence of iron-peak elements in the highest velocity layer ($>0.3c$) with a mass fraction of $\sim 1\%$.  
These new findings pose a challenge of how to transport a considerable amount of explosive nucleosynthesis products, which are produced in the deep layer of the star, into the outermost layer. 
Although the jet activity and the associated chemical mixing in the SN ejecta is a promising hypothesis, it has not been quantitatively investigated by using numerical simulations. 

The role of a jet or a bipolar outflow in the explosive nucleosynthesis has been paid great attention \citep{2002ApJ...565..405M,2003ApJ...598.1163M,2007ApJ...657L..77T}. 
The characteristic nucleosynthesis products potentially distinguish the different jet propagation regimes and/or the central compact object. 
The production of $^{56}$Ni is particularly important since it powers the optical emission of stripped-envelope SNe. 
The observationally derived $^{56}$Ni mass for GRB-SNe is larger than those for normal CCSNe and in the range of $0.1$--$1\,M_\odot$ \citep[e.g.,][]{2017AdAst2017E...5C}. 
Synthesizing a sufficient amount of $^{56}$Ni is, therefore, a requirement for any successful scenario for GRB-SNe. 
Recently, \cite{2018ApJ...860...38B} have performed hydrodynamic simulations of a jet in an analytic progenitor star with a simple $^{56}$Ni production criterion. 
They also conducted post-process radiative transfer calculations for the direct comparison between their synthetic spectra and observed SNe Ic-BL. 
There have also been more recent numerical studies with parametrized jet models motivated by central engine simulations \citep{2021MNRAS.501.2764G,2021arXiv210508092S}. 
However, what powers the SN and jet components, and how the explosive nucleosynthesis proceeds are still debated. 

The jet propagation in a massive star with different settings is important for understanding the dynamical properties of the GRB jet and has been investigated by phenomenological simulation studies in 2D \citep{2000ApJ...531L.119A,2005ApJ...629..903L,2007ApJ...665..569M,2011ApJ...731...80N,2012ApJ...746..122D,2012ApJ...751...57D,2013ApJ...777..162M,2015ApJ...806..205D,2017MNRAS.469.2361H,2018MNRAS.478.4553D} and 3D \citep{2008ApJS..176..467W,2013ApJ...767...19L}, and in the presence of magnetic fields \citep[e.g.,][]{2010NewA...15..749T,2016MNRAS.456.1739B,2020MNRAS.498.3320G,2021arXiv210914619G}. 
These numerical studies usually assume a long-lasting engine required for explaining the ultra-relativistic jet and its prompt gamma-ray emission. 
In terms of unifying low-luminosity GRBs associated with nearby GRB-SNe and luminous GRBs at cosmological distances, the jet injection condition (jet power, duration, opening angle, and so on) can be an important ingredient \citep{2006ApJ...651..960M,2012ApJ...750...68L}. 
A weak or short-duration jet fails to penetrate the star and produce sub-relativistic quasi-spherical ejecta rather than an ultra-relativistic jet. 
\cite{2011ApJ...740..100B} investigated a self-similar expansion law for the jet in a stratified medium and derived the jet penetration condition.

The jet-star interaction may not be the only factor shaping the jet and the cocoon. 
The progenitor environment is also a potential ingredient to determine jet dynamics. 
While it is highly likely that the prompt gamma-ray emission and the afterglow from cosmological GRBs are produced by an ultra-relativistic jet emanating from the progenitor star, the less luminous gamma-ray emission from nearby GRBs could be a consequence of a jet failed to accelerate up to ultra-relativistic speeds. 
\cite{2013ApJ...764L..12S} have conducted a series of jet simulations with a dense circum-stellar medium (CSM) surrounding the progenitor star and suggested that the cocoon component associated with the jet can dissipate its expansion kinetic energy by colliding with the dense CSM, which ultimately leads to a prolonged thermal emission similar to the nearby GRB 060218. 
By introducing an even more dense CSM or an extended envelope that can choke the jet, \cite{2015ApJ...807..172N} proposed a scenario unifying low- and high-luminosity GRBs. 
More recently, \cite{2020ApJ...900..193D} used 2D hydrodynamics simulations to investigate a condition for the jet to stop in a CSM. 
The presence of such a dense CSM may have already been supported by observations of GRB-SNe \citep{2015ApJ...805..159M}. 

All these observational and theoretical progresses ever made suggest that the newly available early spectra of GRB-SNe and the appropriate theoretical interpretations can shed new light on the origin of GRB-SNe and warrant thorough investigations. 
In this work, we perform 3D special-relativistic hydrodynamic simulations of the jet propagation in a massive star to clarify the role of the jet in producing high-velocity ejecta and the jet-induced mixing of heavy elements in the ejecta. 
We consider phenomenological jet propagation in three representative settings, a standard GRB jet, a weak jet, and a jet choked by a dense CSM or an extended envelop. 
We show that the density and chemical structures of the outer ejecta of GRB-SNe implied by recent observations are consistent with those obtained in jet simulations. 
Then we claim that the presence of explosive nucleosynthesis products in the outermost layer of GRB-SNe is a natural consequence of the jet penetration through the massive star. 
This paper is organized as follows. 
In Section \ref{sec:numerical_setups}, we describe the setups for our numerical simulations. 
In Section \ref{sec:results}, we present the results of the simulations. 
The simulation results are discussed in Section \ref{sec:discussion} along with their comparisons with observed GRB-SNe. 
Finally, in Section \ref{sec:summary}, we summarize this paper.

\section{Numerical setups}\label{sec:numerical_setups}
\subsection{Numerical code}
We use the same numerical code as our previous studies on SN explosions \citep{2017MNRAS.466.2633S,2019ApJ...880..150S}. 
The code is designed for multi-dimensional special-relativistic hydrodynamics simulations of expanding gas, such as an SN ejecta and a GRB jet. 
An adaptive mesh refinement (AMR; \citealt{1989JCoPh..82...64B}) technique is employed to cover the whole ejecta and small inner structures simultaneously. 
The maximum AMR refinement level, which determines the finest resolution, decreases along with the expansion of the ejecta. 

The code solves a set of special-relativistic hydrodynamics equations in 3D Cartesian coordinates $(x,y,z)$. 
We denote the local density, the three components of the velocity, and the pressure by $\rho$, $v_i$ ($i=x,y$, and $z$), and $P$, respectively. 
We assume an ideal-gas equation of state with an adiabatic index of $\gamma=4/3$, which is appropriate for relativistic flows with the internal energy dominated by relativistic particles and/or radiation. 
Therefore, the internal energy $e$ is related with the local pressure by
\begin{equation}
    e=\frac{P}{\gamma-1}=3P. 
\end{equation}

We consider a simulation box with $-4.096\times 10^{14}\,\mathrm{cm}\leq x,y\leq 4.096\times 10^{14}\,\mathrm{cm}$ and $0\leq z\leq 4.096\times 10^{14}\,\mathrm{cm}$, which is covered by $128\times 128\times 64$ coarsest grids. 
An equatorial symmetry is assumed and thus only the upper half of the system $(z>0)$ is considered. 
Initially, the maximum AMR refinement level is set to $19$, which corresponds to the minimum resolved length $2^{19}$ times smaller than that of the coarsest grids, $\Delta x=1.22\times 10^7\,\mathrm{cm}$. 
Since the jet propagates almost at the speed of light $c$, the simulation box can cover the jet expansion up to $t\simeq 1.3\times 10^4$ s. 
We stop our simulations $10^4$ s after the initiation. 

\subsection{Progenitor model}
We adopt a $16M_\odot$ carbon-oxygen (CO) core (16TI model) created by \cite{2006ApJ...637..914W} as a stripped-envelope progenitor star model. 
This progenitor model is widely used in long GRB jet studies. 
And the highly energetic explosion of such a massive CO core is known to reproduce the light curve and spectral properties of the prototypical GRB-SN 1998bw \citep[e.g.,][]{1998Natur.395..672I}. 
Therefore, it is a reasonable choice to study this progenitor model as a first step, although possible progenitor dependence should be explored in future studies. 

The progenitor star is surrounded by ambient gas having been produced before the iron-core collapse as a stellar wind or massive mass ejection. 
As we shall introduce in the following subsection, we consider models with and without a dense CSM. 
In cases without a dense CSM, we simply assume a steady stellar wind with the radial density profile given by,
\begin{equation}
    \rho_\mathrm{w}(r)=\frac{\dot{M}}{4\pi v_\mathrm{w} r^2},
\end{equation}
where $r=(x^2+y^2+z^2)^{1/2}$ and $\dot{M}$ and $v_\mathrm{w}$ are constant wind mass-loss rate and velocity. 
We set $\dot{M}/(4\pi v_\mathrm{w})=5\times 10^{11}\,\mathrm{g}\,\mathrm{cm}^{-1}$ (i.e., $\dot{M}\simeq 10^{-5}\,M_\odot\,\mathrm{yr}^{-1}$ for $v_\mathrm{w}=10^3\,\mathrm{km}\,\mathrm{s}^{-1}$).  
This gives only a negligible wind mass for the jet propagation within the simulation box (less than $10^{-5}\,M_\odot$ within $r=3\times 10^{14}\,\mathrm{cm}$).  

We also consider a dense spherical CSM whose radial density profile is expressed as,
\begin{equation}
    \rho_\mathrm{csm}(r)=\frac{M_\mathrm{csm}}{4\pi r^2R_\mathrm{csm}\Gamma(1/p)}\exp\left[-\left(\frac{r}{R_\mathrm{csm}}\right)^p\right],
\end{equation}
where $\Gamma(x)$ is a gamma function and $p$ is set to $p=10$. 
The CSM is extended to $r=R_\mathrm{csm}$, beyond which the CSM density profile shows a sharp drop. 
The circumstellar environments in the close vicinity of GRB progenitors are highly uncertain. 
Although a wide parameter survey is required to clarify how the jet dynamic is influenced by the CSM mass and radius, we try a single set of these parameters, $M_\mathrm{csm}=0.1M_\odot$ and $R_\mathrm{csm}=2.8\times 10^{13}$ cm, which turns out to represent a CSM sufficiently massive to choke the GRB jet. 
For a model with the dense CSM, the ambient density profile is therefore given by the sum of the wind and CSM components, $\rho_\mathrm{w}(r)+\rho_\mathrm{csm}(r)$. 

\subsection{Energy injection}
Since the physical mechanism responsible for the highly energetic explosion and the jet formation in GRB-SNe is poorly understood, we employ phenomenological jet and energy injection procedures similar to previous works.

\subsubsection{Thermal bomb injection}
Most jet simulations assume a constant energy injection rate of the order of $ 10^{50}$--$10^{51}\,\mathrm{erg}\,\mathrm{s}^{-1}$ for several $10\,\mathrm{s}$ to reproduce an ultra-relativistic jet. 
However, such a gradual energy injection cannot produce enough amount of $^{56}$Ni to power the associated SN light \citep[e.g.,][]{2015MNRAS.451..282S,2017ApJ...839...85C}. 
For example, \cite{2018ApJ...860...38B} assumed a time-dependent energy injection with the initial rate as high as $10^{51}\,\mathrm{erg}\,\mathrm{s}^{-1}$ to overcome the Ni deficit. 
We instead assume a two-step engine, the spherical thermal bomb and the collimated jet injection. 
We suppose that an explosion energy of $10^{52}\,\mathrm{erg}$ is initially deposited instantaneously and then the jet injection for a limited solid angle immediately follows.

The initial instantaneous energy deposition is called the thermal bomb. 
The deposited energy for the entire star is assumed to be $E_\mathrm{sn}=10^{52}\,\mathrm{erg}$ (i.e., $E_\mathrm{sn}/2=5\times 10^{51}\,\mathrm{erg}$ is deposited in the simulation box covering the upper hemisphere). 
In practice, we assume an energy deposition in a spherical manner within the injection radius of $R_\mathrm{in}=10^9\,\mathrm{cm}$, below which the star is excised. 
The injection radius encloses the inner core of $\simeq 2.5\,M_\odot$. 
We also note that the thermal bomb injection is required for the consistency with the assumed chemical abundance in the ejecta (Section \ref{sec:chemical}) but has a limited impact on the formation of the high-velocity ejecta because the quasi-spherical blast wave by the thermal bomb is left behind the jet. 

\subsubsection{Jet injection}
The jet injection condition is specified by the following parameters; the luminosity $L_\mathrm{jet}$, the duration $t_\mathrm{jet}$, the half opening angle $\theta_\mathrm{jet}$, the initial Lorentz factor $\Gamma_\mathrm{0}$, and the initial specific energy $\epsilon_\mathrm{0}$. 
The injection radius is the same as the thermal bomb radius, $r=R_\mathrm{in}$. 
In practice, we assume a monotonic increase at a constant rate for each of the mass $M$, momentum $S_i$ ($i=x,y,z$), and energy $E$ of the gas in the conical region with $r\leq R_\mathrm{in}$ and $z/r\geq\cos\theta_\mathrm{jet}$ until $t=t_\mathrm{jet}$. 
The corresponding increases in $M$, $S_i$, and $E$ per unit time are described as,
\begin{equation}
    \dot{M}=
    \frac{L_\mathrm{jet}}{\Gamma_0(c^2+\gamma \epsilon_0)}
\end{equation}
\begin{equation}
    \dot{S}_i=
    \frac{L_\mathrm{jet}v_\mathrm{jet}}{c^2}n_i,
\end{equation}
($n_x=x/r$, $n_y=y/r$, and $n_z=z/r$) and 
\begin{equation}
    \dot{E}=
    L_\mathrm{jet},
\end{equation}
with
\begin{equation}
    v_\mathrm{jet}=\sqrt{1-\frac{1}{\Gamma_\mathrm{0}^2}},
\end{equation}

We fix the free parameters except for the duration to the following values similar to widely used one; $L_\mathrm{jet}=2.5\times10^{50}\,\mathrm{erg}\,\mathrm{s}^{-1}$, $\theta_\mathrm{jet}=10^\circ$, $\Gamma_\mathrm{0}=5$, and $\epsilon_\mathrm{0}/c^2=20$. 
We perform jet simulations in three different setups; the standard, the weak, and the choked jet models. 
In the standard and choked jet models, we assume the jet duration of $t_\mathrm{jet}=20\,\mathrm{s}$, which is long enough for the jet to penetrate the progenitor star. 
In these models, the injected energy by a jet is $L_\mathrm{jet}t_\mathrm{jet}=5\times 10^{51}\,\mathrm{erg}$ (since we only treat the upper hemisphere, the jet energy should be doubled for the entire star with a pair of the bipolar jets). 
On the other hand, we assume a shorter jet duration of $t_\mathrm{jet}=4\,\mathrm{s}$ for the weak jet model. 
As we shall see below, the time required for the jet penetration of the progenitor is $\simeq 5\,\mathrm{s}$. 
Therefore, the weak jet model leads to a jet stuck in the star. 
The jet injected energy in this model is $L_\mathrm{jet}t_\mathrm{jet}=10^{51}\,\mathrm{erg}$ in one hemisphere.

\subsection{Chemical abundance\label{sec:chemical}}
The chemical abundance of the ejecta is calculated as follows. 
Because of the large numerical cost of 3D simulations, it is not practical to carry out full nucleosynthesis calculations along with the simulations or track the abundances of all the elements in the progenitor model. 
We instead specify concentric layers with characteristic elements supposedly produced by the thermal bomb injection. 
Several studies have already investigated the explosive nucleosynthesis operating in the spherical explosion of massive stars with an explosion energy of $10^{52}$ erg \citep[e.g.,][]{1998Natur.395..672I,2000ApJ...534..660I,2001ApJ...550..991N}. 
These studies found that the explosive nucleosynthesis produces concentric layers with characteristic nuclei. 
The innermost layer with $\simeq 0.4\,M_\odot$ is abundant in iron-peak elements, especially $^{56}$Ni, produced by complete Si burning. 
This innermost layer is surrounded by $\simeq 1.0\,M_\odot$ of materials having experienced incomplete Si burning. 
An O burning layer having $\sim 2.0\,M_\odot$ is located outside the two inner layers. 
Then the remaining outer envelope is predominantly composed of unburned C and O. 
Being motivated by the results of these studies, we consider the three layers (layers 1, 2, and 3 from inner to outer) outside the injection radius $r=R_\mathrm{in}$. 
The mass of each layer is $0.4$, $1.0$, and $2.0\,M_\odot$ for the layer 1, 2, and 3, respectively. 
We introduce the mass fractions $X_1$, $X_2$, and $X_3$ of materials initially in the layers 1, 2, and 3 and set the initial value of $X_1$, $X_2$, and $X_3$ as follows,
\begin{eqnarray}
&&(X_1,X_2,X_3)
\nonumber\\
&&=\left\{
\begin{array}{ccl}
(1,0,0)&\mathrm{for}&0\leq M(r)-M(R_\mathrm{in})\leq 0.4M_\odot,\\
(0,1,0)&\mathrm{for}&0.4M_\odot\leq M(r)-M(R_\mathrm{in})\leq 1.4M_\odot\\
(0,0,1)&\mathrm{for}&1.4M_\odot\leq M(r)-M(R_\mathrm{in})\leq 3.4M_\odot\\
\end{array}
\right.,
\end{eqnarray}
where $M(r)$ is the mass enclosed within a given radius $r$ and $M(R_\mathrm{in})\simeq 2.5\,M_\odot$ for the adopted progenitor model. 
We also define the mass fraction $X_\mathrm{ej}$ of the entire ejecta to distinguish the ejecta materials from the CSM;
\begin{equation}
    X_\mathrm{ej}=\left\{
    \begin{array}{l}
    1.0\ \ \ \mathrm{for}\ 0\leq M(r)-M(R_\mathrm{in})\leq 11.4M_\odot,
    \\
    0\ \ \ \mathrm{otherwise.}
    \end{array}\right.
\end{equation}
The mass fraction distributions are evolved as passive scalars along with the hydrodynamic variables. 
In the following, we simply refer materials initially located in the three layers to the iron-peak elements, the incomplete Si burning products, and the O burning products, from inner to outer.

\section{Results}\label{sec:results}
In this section, we present the results of the jet propagation simulations. 
First, we focus on the standard jet model in Section \ref{sec:standard_jet}. 
Then, we present the results of the weak and choked jet models with particular focuses on their differences from the standard model in Sections \ref{sec:weak_jet} and \ref{sec:choked_jet}, respectively. 
The mass and energy spectra of the three jet models are compared in Section \ref{sec:spec}. 

\subsection{Standard jet model}\label{sec:standard_jet}
\begin{figure*}
\begin{center}
\includegraphics[scale=0.48]{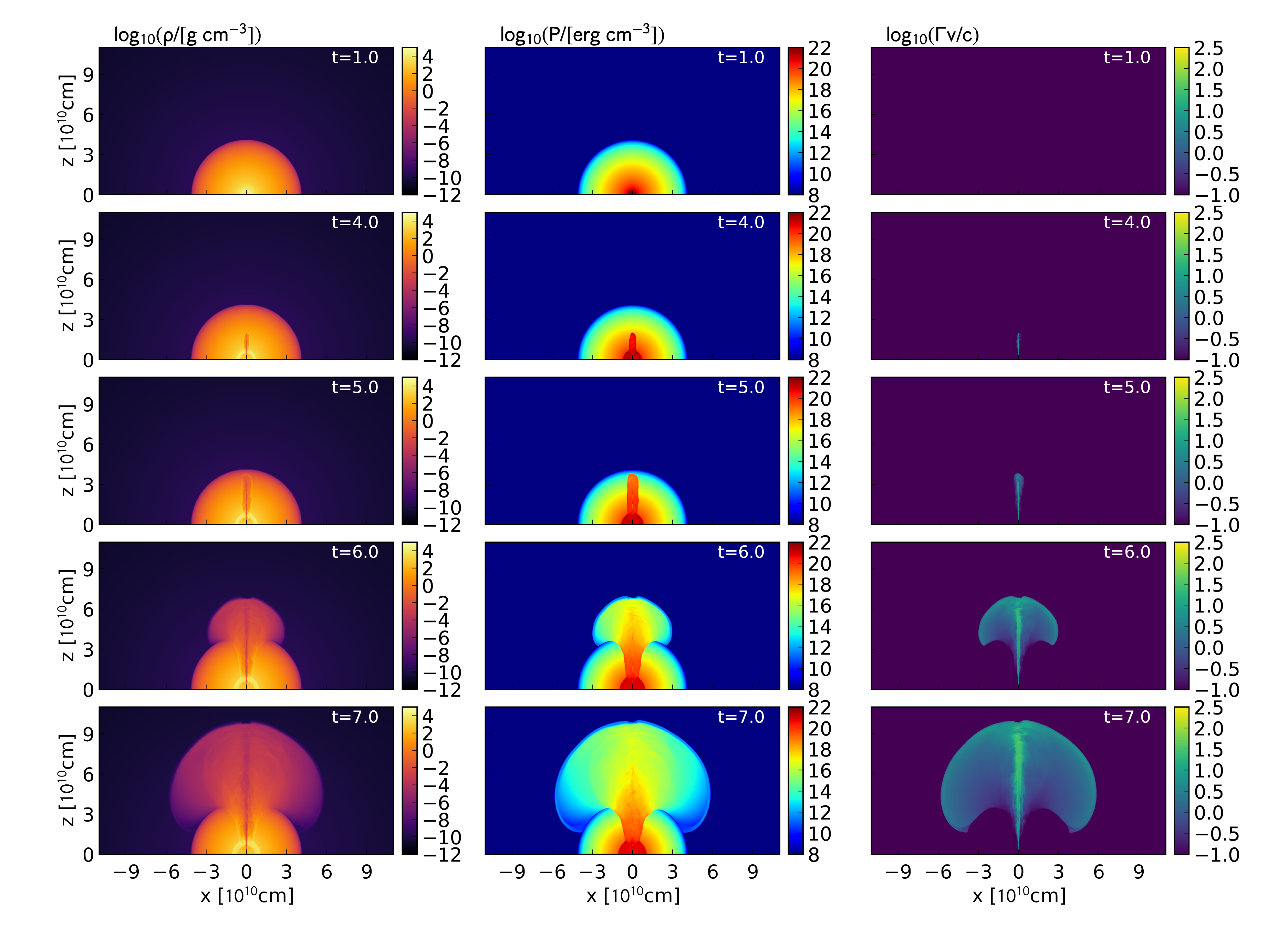}
\caption{Jet propagation in the standard jet model. 
The $x$--$z$ slices ($y=0$) of the color-coded spatial distributions of the density (left), the pressure (center), the 4-velocity (right) are shown. 
Each column represents the distributions at $t=1.0$, $4.0$, $5.0$, $6.0$, and $7.0\,\mathrm{s}$ from top to bottom. }
\label{fig:evol_fiducial}
\end{center}
\end{figure*}
\begin{figure*}
\begin{center}
\includegraphics[scale=0.48]{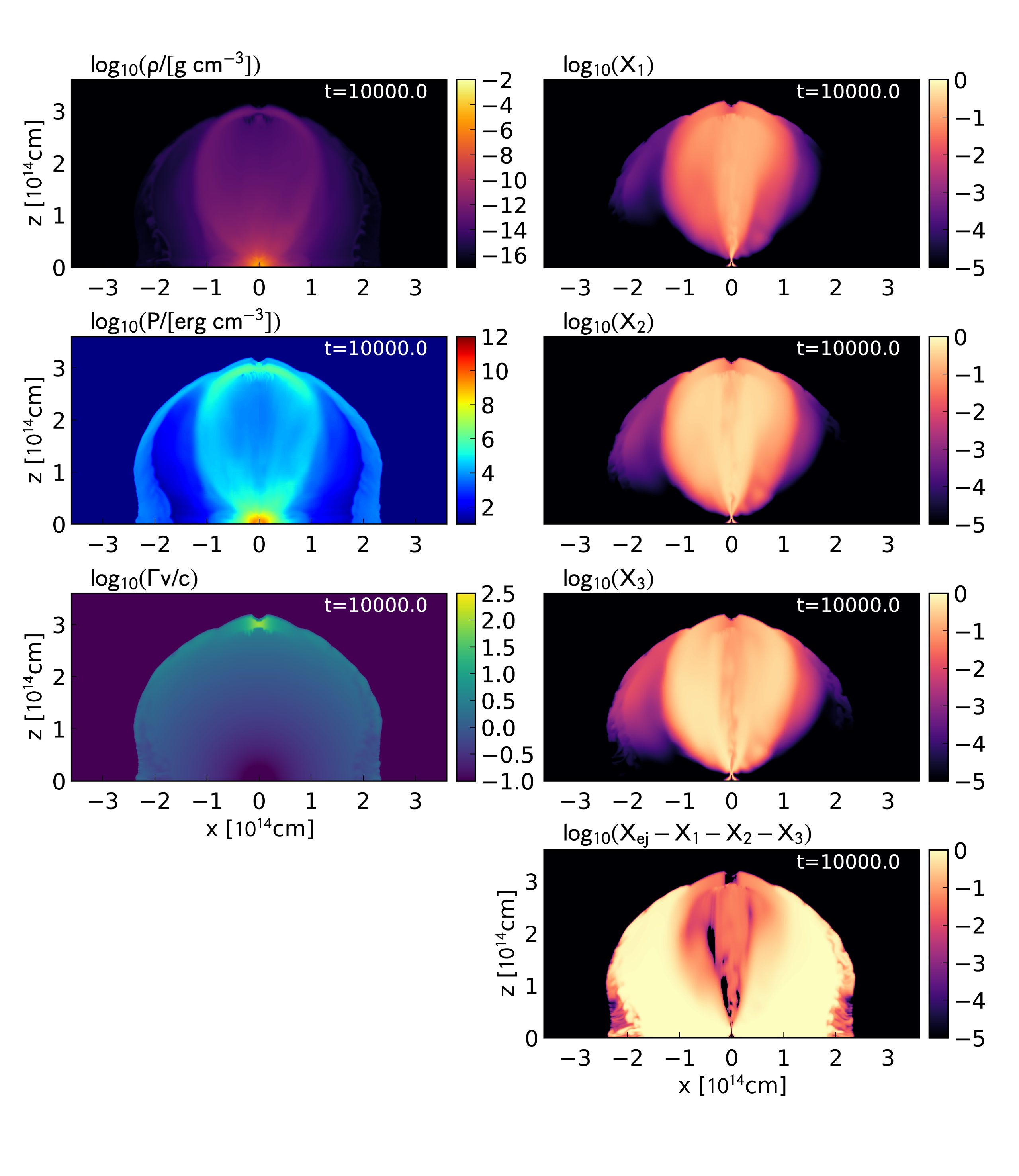}
\caption{Color-coded spatial distributions of physical variables at $t=10^4\,\mathrm{s}$ on the $x$--$z$ plane ($y=0$). 
The left panels show the distributions of the density (top), the pressure (middle), and the 4-velocity (bottom). 
The right panels show the mass fraction distributions of the inner layers $X_1$, $X_2$, and $X_3$, and the remaining part of the ejecta $X_\mathrm{ej}-X_1-X_2-X_3$ from top to bottom.}
\label{fig:final_fiducial}
\end{center}
\end{figure*}
\subsubsection{Jet propagation}
In the standard jet model, the jet dynamics inside and outside the star and the associated cocoon formation are consistent with previous simulation studies. 
Figure \ref{fig:evol_fiducial} illustrates the jet propagation before and after the jet emergence from the progenitor star. 
The energy injection creates a hot ``fireball'' with an outgoing momentum within the injection radius.  
The expanding jet initially experiences the ram pressure of the stratified material as it moves forward. 
The reconfinment shock then develops and makes the jet well collimated in the star.
The collimated jet drills through the star, leaving  a hole along the jet axis. 
The jet takes about $5\,\mathrm{s}$ (see, the middle panels of Figure \ref{fig:evol_fiducial}) to reach the stellar surface. 
After the jet emergence, the stellar material pushed by the jet head starts expanding in the lateral direction. 
Since the jet injection is still ongoing after $t=5\,\mathrm{s}$, the jet head is followed by a high Lorentz factor flow as seen in the bottom panels of Figure \ref{fig:evol_fiducial}. 
The laterally expanding material forms the so-called cocoon, which expands in a quasi-spherical way. 
The jet properties, such as the breakout time and the morphology of the high-Lorentz factor flow, in our simulation are reassuringly similar to previous 3D jet simulations \citep[e.g.,][]{2013ApJ...767...19L} with similar numerical setups (the jet injection condition and the progenitor model). 

\begin{figure*}
\begin{center}
\includegraphics[scale=0.55]{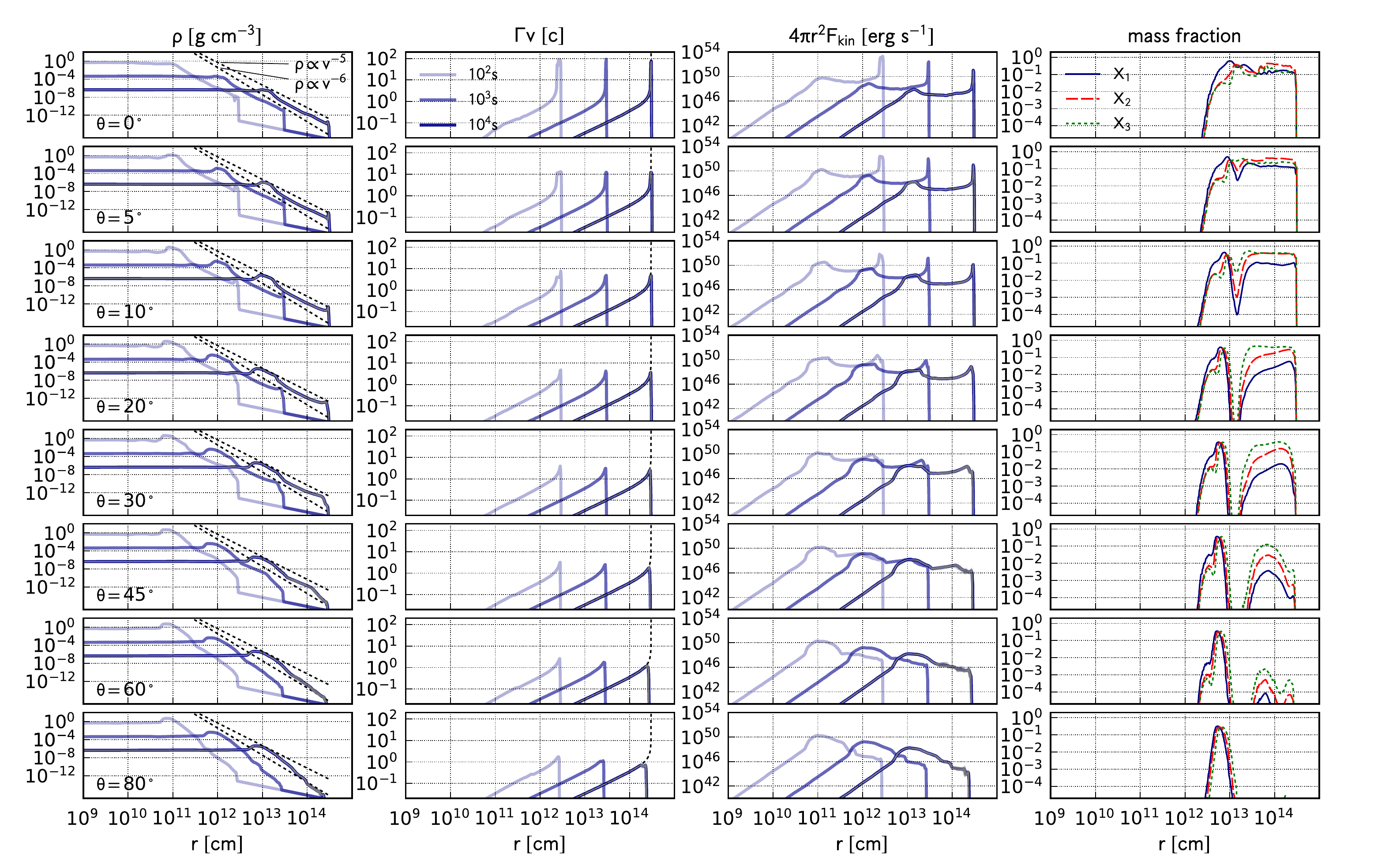}
\caption{Azimuth angle-averaged radial profiles for different inclination angles $\theta=0^\circ$, $5^\circ$, $10^\circ$, $20^\circ$, $30^\circ$, $45^\circ$, $60^\circ$, and $80^\circ$ from top to bottom. 
Each column represents the distributions of the density, the 4-velocity, the outgoing kinetic power, and the mass fractions of the inner layers from left to right (blue thick lines). 
{\it Left three columns}; the distributions at $t=10^2$ $10^3$, and $10^4$ s are plotted. 
For the distributions at $t=10^4$ s, profiles along $18$ different azimuth angles are also plotted as gray thin lines to highlight the variety of the distributions. 
{\it Rightmost column}; the mass fraction profiles are plotted only at $t=10^4$ s. }
\label{fig:radial_fiducial}
\end{center}
\end{figure*}

After the jet termination at $t=20\,\mathrm{s}$, the high-Lorentz factor flow leaves the injection region. 
The jet material injected after the jet emergence almost freely travels in the hole along the jet axis. 
The freely traveling jet continues to accelerate from the initial Lorentz factor $\Gamma_0=5$ up to $\Gamma\simeq 100$ by converting its initial internal energy into the kinetic one. 
Since the ambient material surrounding the progenitor star is sufficiently dilute, the high-Lorentz factor bullet keeps its high Lorentz factor even at the end of the simulation ($t=10^4\,\mathrm{s}$). 
The spatial distributions of physical variables at $t=10^4\,\mathrm{s}$ are shown in Figure \ref{fig:final_fiducial}. 
The density structure shows the elongated stellar material, the high-Lorentz factor bullet on the top, and the quasi-spherical cocoon moving at mildly relativistic speeds. 
The mass fraction distributions of the three inner layers are also shown in Figure \ref{fig:final_fiducial} (right panels). 
The distributions of $X_1$, $X_2$, and $X_3$ share a similar shape, while the average mass fractions are different from each other. 
The mass fraction distributions elongated along the jet axis clearly indicate that the jet propagation has played an important role in transporting and mixing nucleosynthesis products into the outermost layers.

\subsubsection{Radial profiles}
We obtain a more quantitative look at the ejecta structure in the radial profiles of physical variables with different inclination angles in Figure \ref{fig:radial_fiducial}. 
The set of inclination and azimuth angles, $\theta\equiv\mathrm{acos}(z/r)$ and $\phi\equiv \mathrm{atan}(y/x)$, specifies a ray from the origin $(x,y,z)=(0,0,0)$ in the simulation box, along which we obtain the radial distributions of hydrodynamic variables, e.g., $Q(r,\theta,\phi)$, from the simulation results. 
We then average the profiles on multiple rays with the same inclination angle over $\phi$ to obtain the azimuth angle-averaged radial profiles for the inclination angle $\theta$;
\begin{equation}
    \bar Q(r,\theta)=\frac{1}{2\pi}\int ^{2\pi}_0Q(r,\theta,\phi)d\phi.
\end{equation}
In Figure \ref{fig:radial_fiducial}, we plot the radial profiles of the density, the 4-velocity, the isotropic outgoing kinetic power, and the mass fractions of the layer $1$, $2$, and $3$ for several inclination angles. 
The isotropic outgoing kinetic power $L_\mathrm{kin}$ at $r$ is defined as the product of the spherical surface area $4\pi r^2$ and the kinetic flux $F_\mathrm{kin}$
\begin{equation}
    L_\mathrm{kin}=4\pi r^2F_\mathrm{kin}=4\pi r^2 \rho\Gamma(\Gamma-1)c^2v,
\end{equation}
where $v=(v_x^2+v_y^2+v_z^2)^{1/2}\equiv \beta c$. 

The 4-velocity profiles along different directions are well reproduced by the free expansion law $v=r/t$ at all the three epochs in Figure \ref{fig:radial_fiducial}, indicating that the ejecta has already entered the free expansion stage at $t>100\,\mathrm{s}$. 
The 4-velocity profiles along $\theta\leq 5^\circ$ are characterized by the jet component traveling at highly relativistic velocities ($\Gamma \beta>10$). 
Correspondingly, the kinetic power profile has a sharp peak around $r\simeq ct$. 
This highly relativistic jet component supposedly dissipates its energy to produce the prompt gamma-ray emission and then the afterglow.  
Even at larger inclination angles of  $\theta\geq 10^\circ$, the ejecta has the outermost layers traveling at relativistic speeds ($\Gamma \beta>1$). 
A considerable amount of the injected energy is thus transported to the outflow at a wide opening angle with the outgoing kinetic power of $10^{46}$--$10^{47}\,\mathrm{erg}\,\mathrm{s}^{-1}$ at the final epoch $t=10^4\,\mathrm{s}$. 

The radial density distributions of the outer sub-relativistic ejecta are well approximated by a simple function of the radius. 
As shown in Figure \ref{fig:radial_fiducial}, they appear to follow a power-law function of the radius (or the radial velocity because of $v=r/t$). 
A power-law function with an exponent of $-5$ or $-6$ well reproduces the outermost density profiles. 
The radial density profiles around the jet axis are more likely to follow $\rho\propto v^{-5}$, while those at larger inclination angles follow $\rho\propto v^{-6}$ or even steeper profiles. 
The innermost part of the ejecta is represented by a flat density profile. 
Then, the inner and outermost parts are connected by the intermediate part with the radial density slope steeper than $\rho\propto v^{-5}$ or $v^{-6}$. 
We note that the radial density profiles realized in spherical SN explosions exhibit a much steeper slope with $d\ln\rho/d\ln r\simeq -10$ \citep{1999ApJ...510..379M}. 
 \cite{2001ApJ...550..991N} have found an outer density slope of $d\ln\rho/d\ln r\simeq -8$ for their energetic explosion models of a massive CO star (a progenitor model for SN 1998bw), but the outer density slope in our jet model is shallower than theirs. 

The radial profiles of the mass fractions $X_1$, $X_2$, and $X_3$ in Figure \ref{fig:radial_fiducial} show the impact of the jet-induced chemical mixing. 
Around the jet axis, the iron-peak elements are dredged up by the jet. 
The mass fraction of the iron-peak elements is as large as $X_1\simeq 0.1$ at the outermost layer with $\theta\leq 10^\circ$. 
The sub-relativistic cocoon along $10^\circ\leq\theta\leq 30^\circ$ also contains at least $1\%$ of the iron-peak elements. 
On the other hand, the radial mass fraction profiles at large inclination angles show a peak at the same radius as the density peak corresponding to the innermost layer with a much smaller radial velocity. 
Therefore, far from the jet axis, a considerable fraction of the iron-peak elements at large inclination angles are still embedded in the slowly expanding SN ejecta. 
This is also seen in the spatial distributions of $X_1$ and $X_2$ in Figure \ref{fig:final_fiducial}.

\subsection{Weak jet model}\label{sec:weak_jet}
\begin{figure*}
\begin{center}
\includegraphics[scale=0.48]{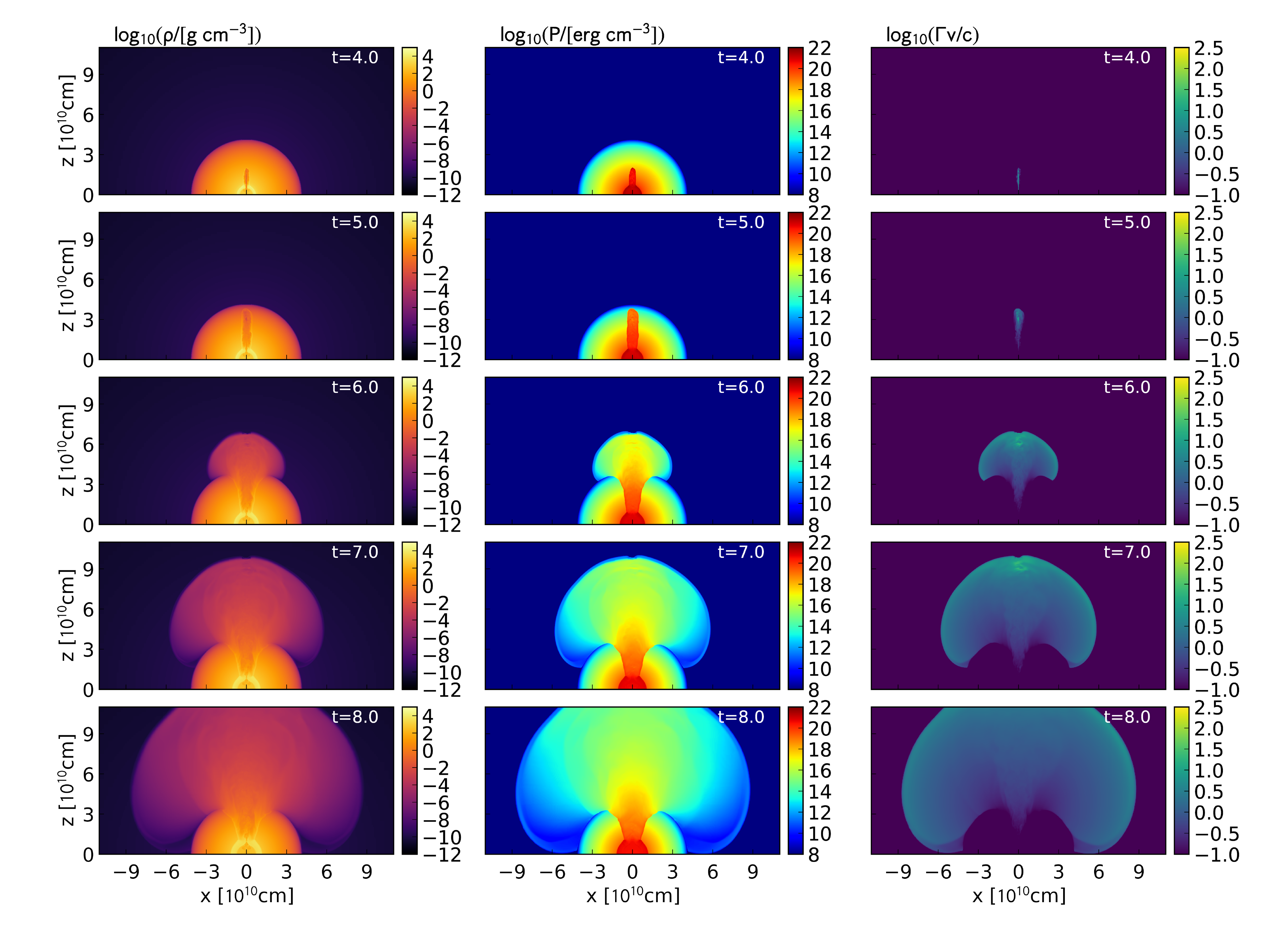}
\caption{Same as Figure \ref{fig:evol_fiducial}, but for the weak jet model at $t=4.0$, $5.0$, $6.0$, $7.0$, and $8.0\,\mathrm{s}$.}
\label{fig:evol_weak}
\end{center}
\end{figure*}
In the weak jet model, the jet is injected only for the short period of $t_\mathrm{jet}=4\,\mathrm{s}$ with the same injection rate as the standard jet model. 
In other words, the jet is terminated while it is still propagating in the progenitor star. 
The resultant dynamical evolution of the ejected material is highlighted in Figure \ref{fig:evol_weak}. 
The quasi-spherical cocoon component is also created in the weak jet model. 
This component is predominantly composed of the stellar envelope pushed by the jet head. 
The cocoon accelerates by consuming its high internal energy, which originates from the dissipation of the jet kinetic energy in the star. 
As a result, the outermost layers reach mildly relativistic speeds. 
In fact, the cocoon shows similar dynamical properties to that of the standard jet model. 
On the other hand, the early termination of the jet injection leads to a failure of an ultra-relativistic jet. 
As seen in the 4-velocity distribution in the right panels of Figure \ref{fig:evol_weak}, the high-Lorentz factor region along the jet axis is missing. 
Such an ejecta unlikely produces a GRB with a luminous prompt gamma-ray emission, but a less energetic transient with softer emission properties may be expected. 

\begin{figure*}
\begin{center}
\includegraphics[scale=0.48]{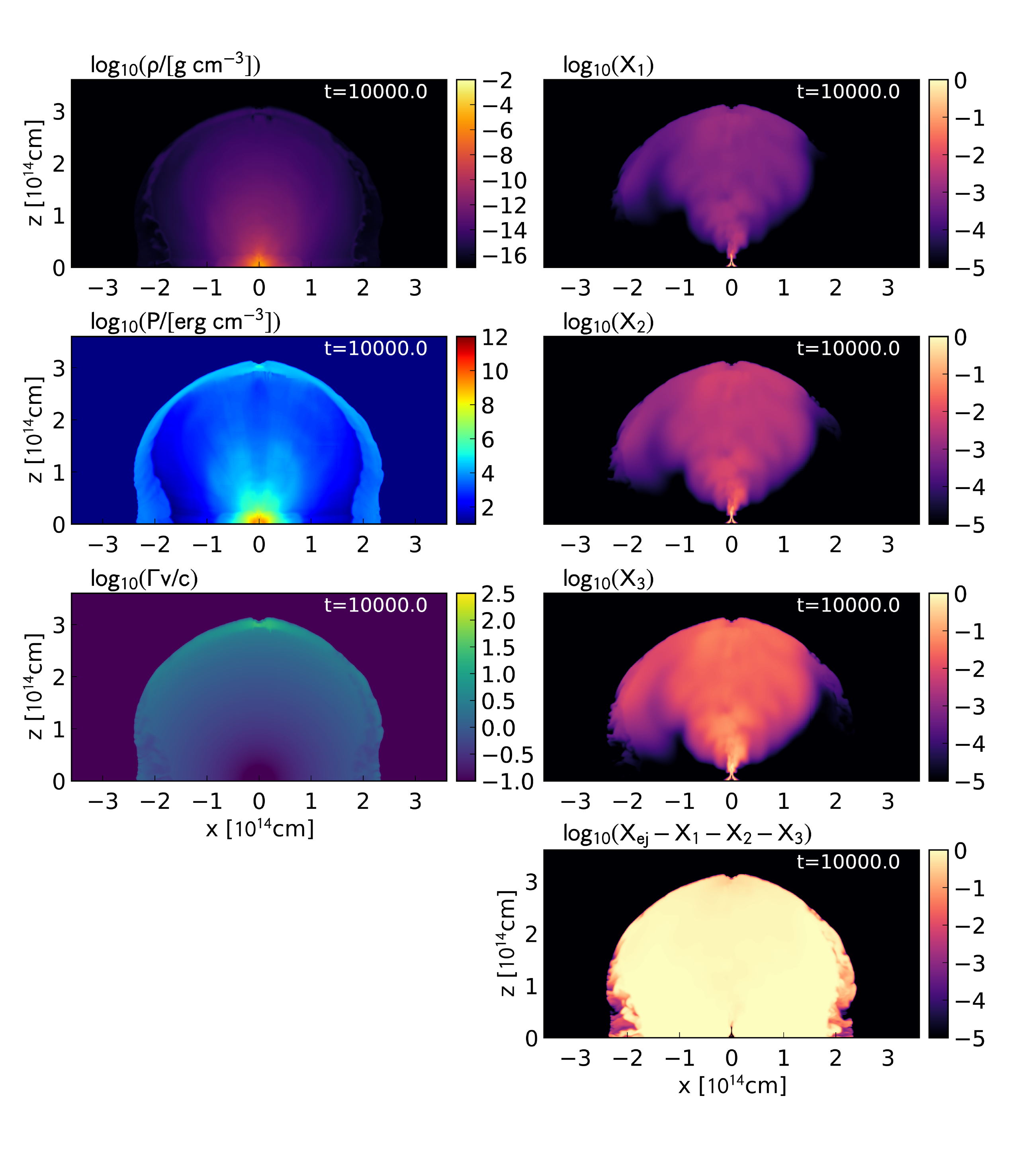}
\caption{Same as Figure \ref{fig:final_fiducial}, but for the weak jet model.}
\label{fig:final_weak}
\end{center}
\end{figure*}

\begin{figure*}
\begin{center}
\includegraphics[scale=0.55]{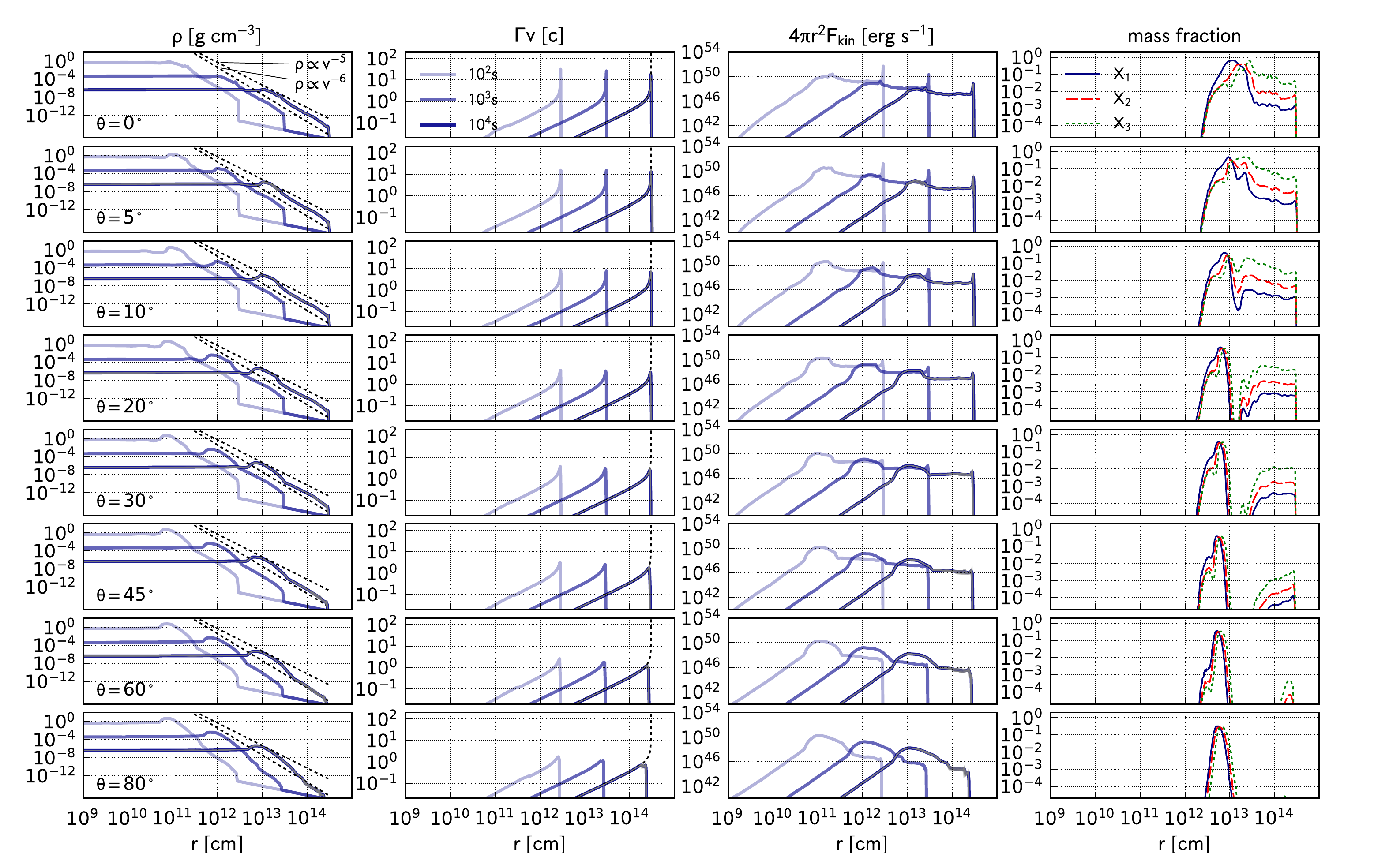}
\caption{Same as Figure \ref{fig:radial_fiducial}, but for the weak jet model.}
\label{fig:radial_weak}
\end{center}
\end{figure*}

Despite its dynamical properties similar to the standard jet model, the cocoon component in the weak jet model exhibits remarkably different elemental distributions.  
The right panels of Figure \ref{fig:final_weak} show the spatial distributions of the mass fractions $X_1$, $X_2$, $X_3$, and the remaining ejecta. 
In contrast to the standard jet model in Figure \ref{fig:final_fiducial}, the mass fraction of the iron-peak elements in the outer ejecta is much smaller by a factor of $\sim 100$, $X_{1}\simeq 10^{-3}$. 
The radial profiles in Figure \ref{fig:radial_weak} also show the mildly relativistic ejecta component much less contaminated by the iron-peak elements and incomplete Si burning products. 
Accordingly, the mass fraction of the unburned material (the right bottom panel of Figure \ref{fig:final_weak}) is close to unity even around the jet axis. 
This remarkable difference between the standard and weak jet models clearly demonstrates the role of the long-lasting jet in mixing inner layers of the star into the outer high-velocity ejecta. 
With the jet-induced chemical mixing, the high-velocity ejecta is highly enriched with heavy elements produced in the explosion and thus the spectral appearance would be different from those expected for the CO layer of a massive star. 

\subsection{Choked jet model}\label{sec:choked_jet}
\begin{figure*}
\begin{center}
\includegraphics[scale=0.48]{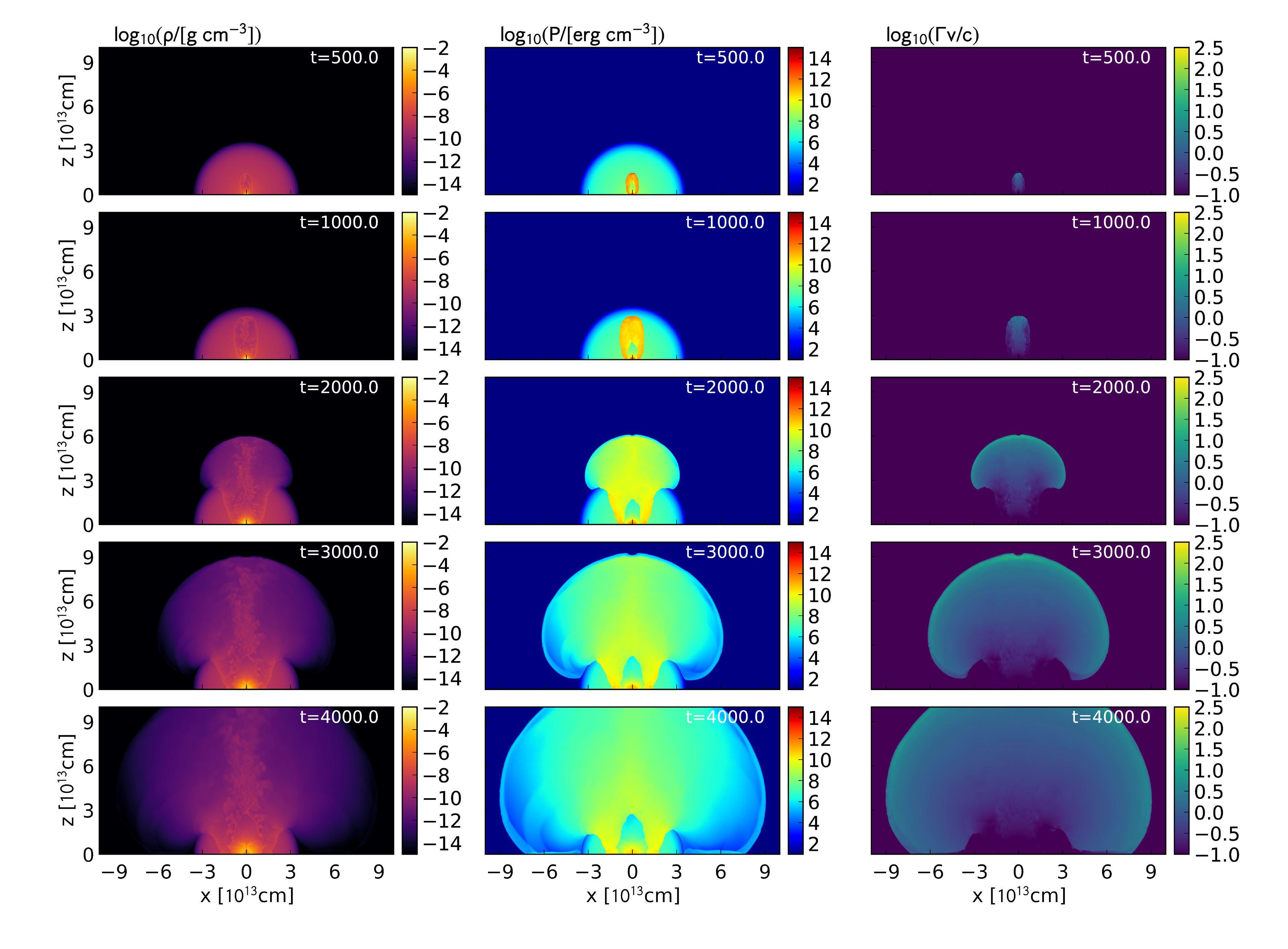}
\caption{Same as Figure \ref{fig:evol_fiducial}, but for the choked jet model at $t=500$, $1000$, $2000$, $3000$, and $4000\,\mathrm{s}$.}
\label{fig:evol_csm}
\end{center}
\end{figure*}

Finally, we consider the choked jet model. 
Figure \ref{fig:evol_csm} shows the dynamical evolution of the jet at later epochs. 
In this model, the extended dense CSM surrounds the progenitor star and thus the jet experiences a high ram pressure even after the emergence from the stellar surface. 
Therefore, the jet is still collimated as long as it propagates in the CSM (the 1st and 2nd rows of Figure \ref{fig:evol_csm}). 
The jet reaches the outer edge of the CSM at $t\simeq 10^3\,\mathrm{s}$ and then emerges from the CSM. 
Since the CSM is massive enough, the jet head has already slowed down to a mildly relativistic speed at the moment of the CSM breakout. 
The ejection of the CSM materials pushed by the jet head happens in a similar way to the jet breakout from the stellar surface in the standard jet model. 
In other words, the breakout is accompanied by the lateral expansion of the accelerated materials to form a quasi-spherical cocoon. 

\begin{figure*}
\begin{center}
\includegraphics[scale=0.48]{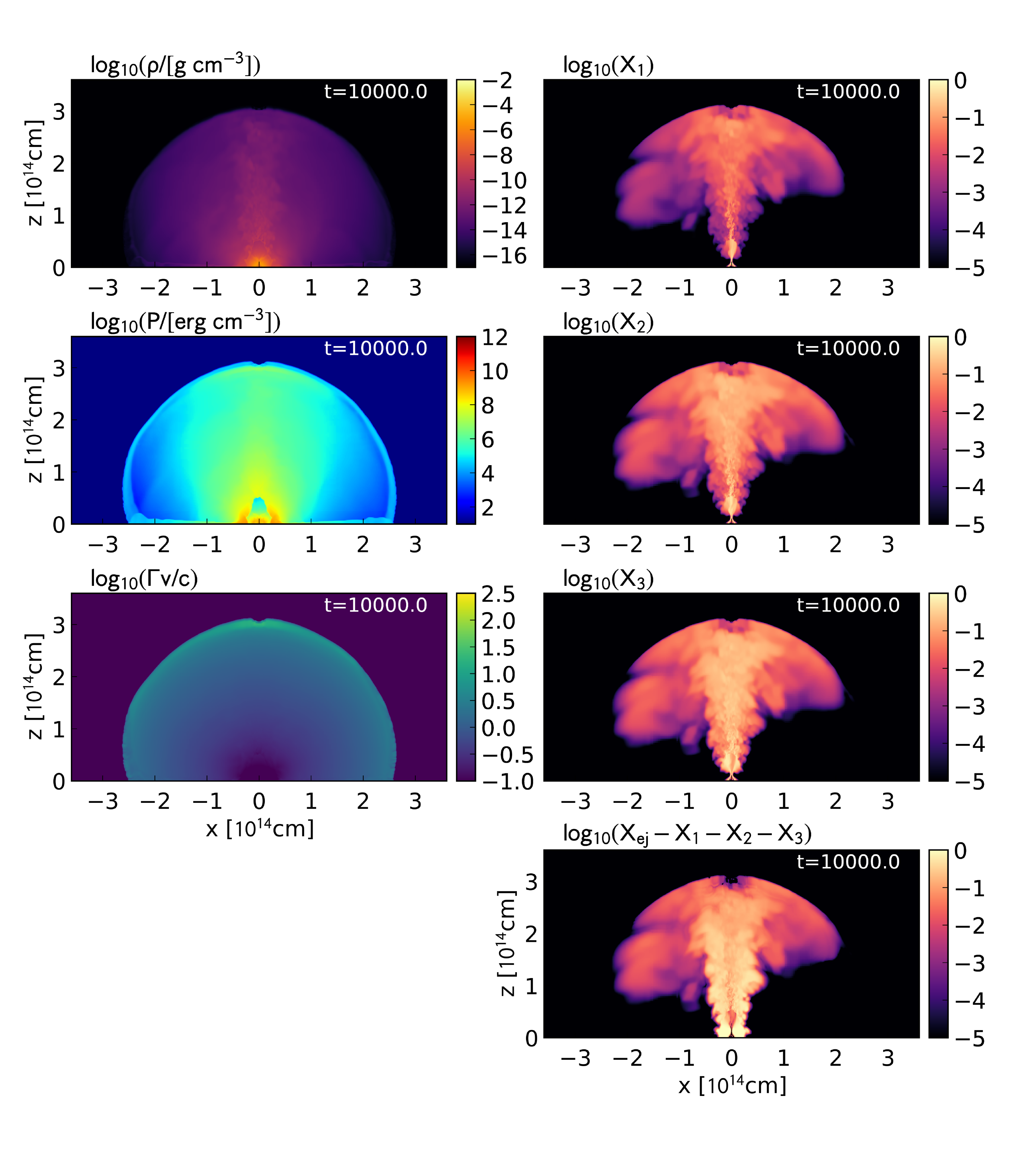}
\caption{Same as Figure \ref{fig:final_fiducial}, but for the choked jet model.}
\label{fig:final_csm}
\end{center}
\end{figure*}

The density, pressure, and 4-velocity distributions at $t=10^4\,\mathrm{s}$ in Figure \ref{fig:final_csm} exhibit several qualitative differences from the other two models. 
The most remarkable difference is found in the pressure distribution. 
The internal energy loaded in the high-velocity ejecta is clearly larger than the other two models at the same epoch. 
This is simply a consequence of the jet-CSM interaction. 
The cocoon components realized in the standard and weak jet models are created soon after the jet breakout and continue to cool in the subsequent expansion stage. 
In the choked jet model, however, the massive CSM effectively decelerates the jet and thus the jet kinetic energy is converted into the internal energy of the shocked gas. 
As a consequence, a much hotter cocoon is realized in this model at a much later epoch than the other two models. 
The jet component, instead, fails to be ultra-relativistic. 
Therefore, this model unlikely produces a luminous GRB. 

The chemical structure also shows a distinct property. 
The right panels of Figure \ref{fig:final_csm} show the different distributions of the iron-peak elements from the standard jet model. 
Since this model assumes the same jet injection condition as the standard jet model, the inner materials are dredged up toward outer envelopes due to the long-lasting jet activity. 
As the jet is collimated in the presence of the dense CSM, the iron-peak elements are also confined in a cylindrical region before the CSM breakout. 
After the breakout, the dredged-up materials are also mixed into the laterally expanding ejecta, leading to the spatial distributions shown in the right panels of Figure \ref{fig:final_csm}. 

\begin{figure*}
\begin{center}
\includegraphics[scale=0.55]{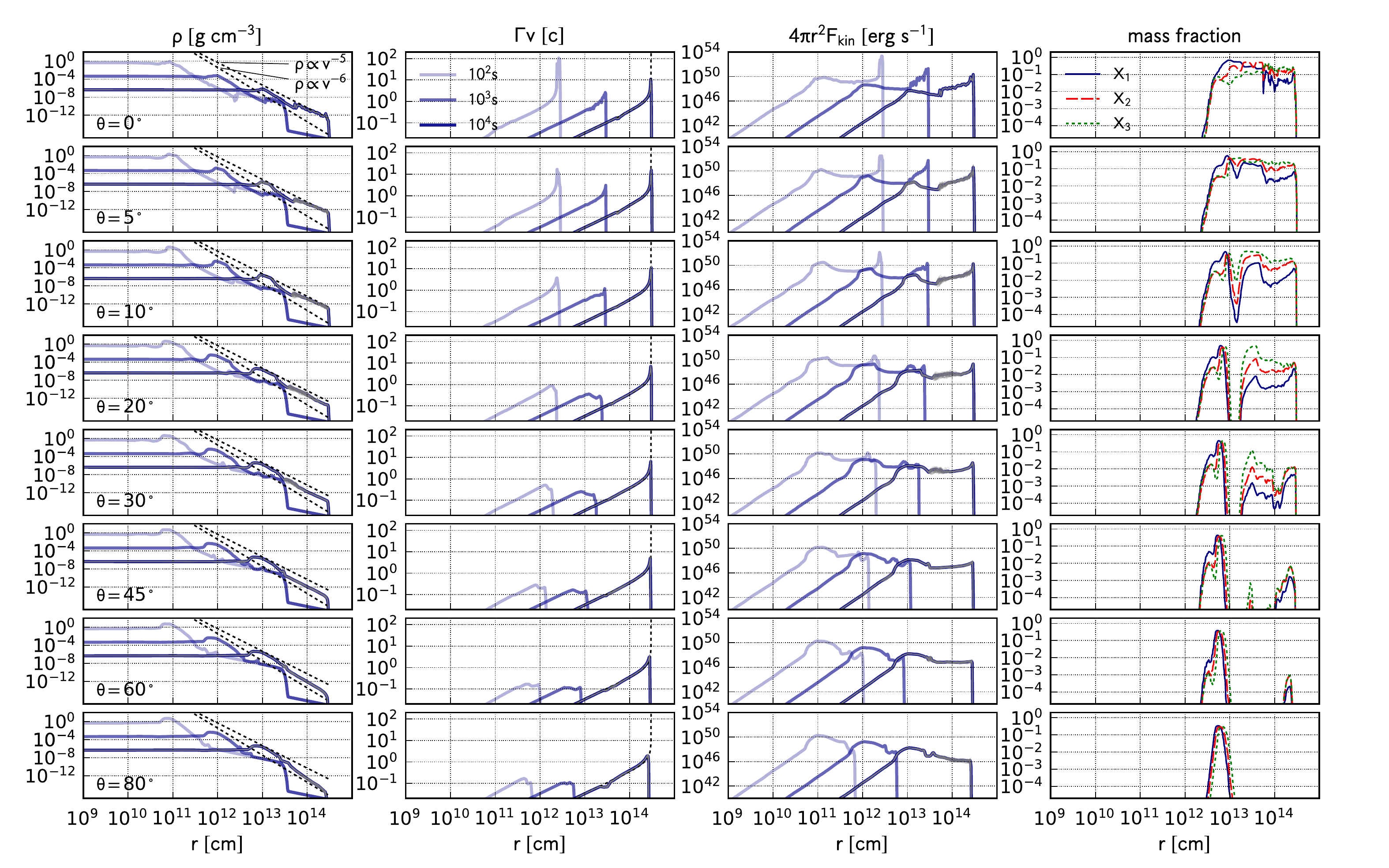}
\caption{Same as Figure \ref{fig:radial_fiducial}, but for the choked jet model.}
\label{fig:radial_csm}
\end{center}
\end{figure*}

The radial profiles of physical variables are shown in Figure \ref{fig:radial_csm}. 
The outer ejecta in this model is dominated by the shocked and accelerated CSM. 
Therefore, the radial density distributions at $t=10^4\,\mathrm{s}$ show relatively massive outer ejecta with a density higher than those of the standard and weak jet models. 
Despite the different total mass, the density profiles in Figure \ref{fig:radial_csm} show outer density slopes similar to $d\ln\rho/d\ln r=-5$, as found in the other two models. 
This is probably because of a similar formation process of the high-velocity quasi-spherical component. 
The jet energy dissipated in the CSM is redistributed within the ejecta traveling into a wider solid angle.  
This results in the high-velocity ejecta with a larger outgoing kinetic power than that in the standard jet model even outside the jet opening angle. 
In contrast to the featureless mass fraction distributions found in Figure \ref{fig:final_fiducial}, those in Figure \ref{fig:final_csm} have shown a more complicated structure. 
While propagating in the CSM, the jet suffers from further mixing of the material with the CSM, resulting in a clumpy structure.

\subsection{Mass and energy spectra\label{sec:spec}}
\begin{figure*}
\begin{center}
\includegraphics[scale=0.55]{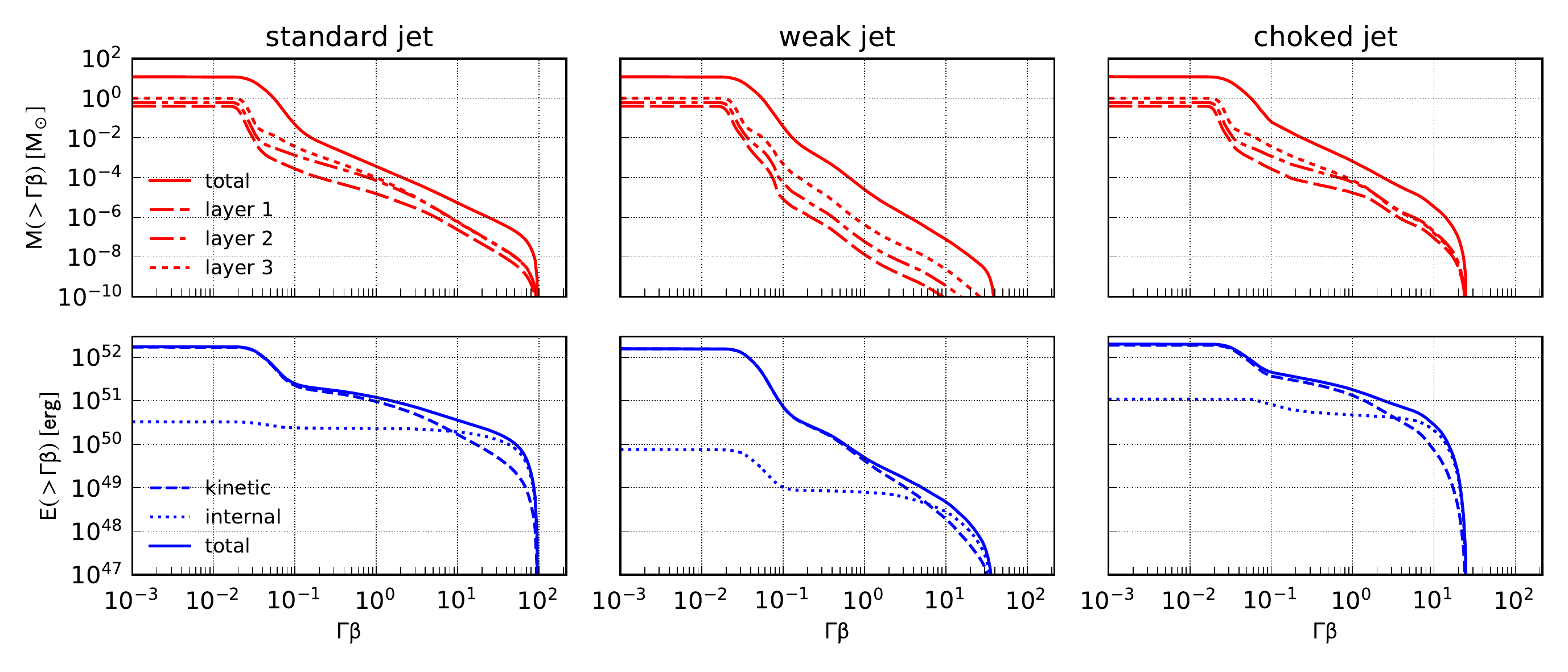}
\caption{Mass and energy spectra of the ejecta for the standard, weak, and choked jet models from left to right. 
Each column presents the mass (upper) and the energy (lower) of the ejecta traveling faster than a threshold 4-velocity. 
In the upper panel, the contributions from the inner layers 1, 2, and 3 are also plotted. 
In the lower panel, the kinetic and internal energies are plotted as dashed and dotted lines, while the thick solid line represents the total energy. }
\label{fig:spec}
\end{center}
\end{figure*}

For further quantitative comparisons between the different models, we calculate the mass and the energy spectra of the ejecta. 
We define the mass and the energy of the ejecta traveling at the 4-velocity faster than a threshold value $\Gamma\beta c$ as follows,
\begin{equation}
    M(>\Gamma\beta)=
    \int_{>\Gamma\beta}\rho\Gamma dV,
    \label{eq:Mspec}
\end{equation}
\begin{equation}
    E_\mathrm{kin}(>\Gamma\beta)=
     \int_{>\Gamma\beta}\rho\Gamma(\Gamma-1) dV,
\end{equation}
for the kinetic energy, and
\begin{equation}
    E_\mathrm{int}(>\Gamma\beta)=
     \int_{>\Gamma\beta}\left(\frac{\gamma}{\gamma-1}\Gamma^2-1\right)PdV,
\end{equation}
for the internal energy, where each integration is carried out only for the numerical cells with the 4-velocity larger than $\Gamma\beta c$. 
In the volume integration, we include the contribution from the lower hemisphere ($z<0$) by assuming the equatorial symmetry. 
In a similar way, we also define the mass spectra for materials having been in the three inner layers 1, 2, and 3 by replacing $\rho$ with $\rho X_i$ ($i=1$, $2$, and $3$) in Equation \ref{eq:Mspec}. 
Figure \ref{fig:spec} compares the mass and the energy spectra for the three jet models. 

All the models show the mass and energy spectra decreasing with the 4-velocity. 
They are composed of the low-velocity ($\Gamma\beta<0.1$) and high-velocity ($\Gamma\beta>0.1$) parts. 
The former shows a flat spectrum, while the latter decreases toward the maximum 4-velocity. 
The standard jet model shows the mass and energy spectra extending to $\Gamma\beta\simeq 100$ due to the successful penetration of the jet and its subsequent propagation through the dilute ambient gas. 
The weak and choked jet models exhibit the maximum 4-velocity of several tens because of the unsuccessful jet penetration through the star and the CSM, respectively. 
The low-velocity parts of the mass and energy spectra are similar among the three models. 
They all show $M(>\Gamma\beta)\simeq 10\,M_\odot$ ejecta with the kinetic energy being $E_\mathrm{kin}(>\Gamma\beta)>10^{52}\,\mathrm{erg}$ for the 4-velocity slower than $0.1c$. 
This is partly because of the isotropic thermal bomb injection, but a part of the jet energy is also dissipated in the star and then contributes to the energy of the low-velocity ejecta. 
A fraction of the jet energy is injected into the high-velocity ejecta with the 4-velocity faster than $0.1c$ to create a high-velocity part extending to even higher $\Gamma \beta$. 

\section{Discussion}\label{sec:discussion}
In this section, we discuss possible electromagnetic signatures of the high-velocity ejecta. 
The thermal/non-thermal emission from a GRB cocoon have already been considered in some previous studies \citep[e.g.,][]{2002MNRAS.337.1349R,2017ApJ...834...28N,2018ApJ...863...32D,2018MNRAS.478.4553D}. 
In this study, we consider the effect of the jet-induced chemical mixing on the early SN spectra and the emission from the high-velocity ejecta interacting with a dense CSM.

\subsection{Early-time spectra as a probe of jet activities}
\subsubsection{Dynamical properties of high-velocity ejecta}
The jet-star interaction produces quasi-spherical ejecta expanding at mildly relativistic speeds. 
In the free expansion stage long after the jet termination, the outermost radial density profiles along different directions follow power-law functions with exponents between $-5$ and $-6$. 
The high-velocity ejecta produced in all the three representative models share similar dynamical properties.  
On the other hand, an instantaneous explosion of a similar progenitor star with the explosion energy of $10^{52}\,\mathrm{erg}$ produces an expanding outer envelope with $\rho\propto v^{-8}$ \citep{2001ApJ...550..991N}. 
Therefore the presence of the high-velocity ejecta with a shallow density slope is considered as a common property among SNe with a long-lasting engine \citep[see also][]{2017MNRAS.466.2633S,2019ApJ...880..150S}. 

The outer density slope of an SN influences its spectral evolution \citep[e.g.,][]{2017MNRAS.469.2498M} and therefore has been paid great attention. 
The spectral evolution of the prototypical GRB-SN 1998bw was modeled in detail by \cite{2001ApJ...550..991N}. 
Although their explosion model exhibited the outer density profile of $\rho\propto v^{-8}$, they found that a slightly shallower profile of $\rho\propto v^{-6}$ above $v=30,000\,\mathrm{km}\,\mathrm{s}^{-1}\simeq 0.1c$ gives a better explanation for the spectral evolution. 
A similar analysis of another energetic Type-Ic SN 1997ef by \cite{2000ApJ...545..407M} has revealed an even shallower slope of $\rho\propto v^{-4}$ at $v>25,000\,\mathrm{km}\,\mathrm{s}^{-1}$. 
More recently, the early-time observations of SN 2017iuk found the spectral evolution consistent with freely expanding spherical ejecta with $\rho\propto v^{-6}$ extending out to $v>0.3c$ \citep{2019Natur.565..324I}. 
These observationally constrained density slopes of energetic SNe are in overall agreement with the shallow density profiles realized in our jet models and therefore suggest the presence of long-lasting engine activity for GRB-SNe.

As we have noted in Section \ref{sec:spec}, in the standard jet model, the mass of the ejecta faster than $\Gamma \beta=0.3$ is of the order of $M(\Gamma\beta>0.3)=10^{-3}\,M_\odot$, while including the slower part ($\Gamma\beta>0.03$) leads to the almost entire progenitor mass (excluding the inner core), $M(\Gamma\beta>0.03)\simeq 10\,M_\odot$, coupled with $E_\mathrm{kin}(\Gamma\beta>0.03)>10^{52}\,\mathrm{erg}$ (Figure \ref{fig:spec}) of the kinetic energy. 
These values are in good agreement with the spectral analysis of SN 2017iuk \citep{2019Natur.565..324I}, reinforcing the idea that the high-velocity component observed in this particular GRB-SN was the cocoon produced by the jet penetration.

\subsubsection{Chemical structure of the high-velocity ejecta}
The chemical composition of the ejecta is also an important ingredient for the spectral evolution of any SN. 
Heavy elements synthesized in explosive nucleosynthesis, particularly iron-peak elements, cause severe UV blocking, making SN spectra redder. 
Therefore, the bluer part of the optical spectra of SNe tells us the heavy metal content in the ejecta. 
Such red spectra are usually found in the late-time observations of SNe, which agrees with the heavy element synthesis in the innermost layer of exploding stars. 
In the early-time observations of GRB 171205A/SN 2017iuk, however,  \cite{2019Natur.565..324I} detected highly blue-shifted absorption features of Ca and Si in the spectra as well as substantial absorption in shorter wavelengths. 
The detailed spectral analysis of SN 2017iuk revealed the high-velocity ejecta with the mass fraction of iron-peak elements being $\sim 0.01$. 
\cite{2019MNRAS.487.5824A} and \cite{2020MNRAS.492.5956A} have reached a similar conclusion for GRB 161219B/SN 2016jca, another GRB-SNe with UV-suppressed early spectra. 
The abundant iron-peak elements in the outermost layer of the GRB-SNe clearly indicate efficient mixing processes operating in the ejecta. 

The mass fraction of iron peak elements at an order of $\simeq 0.01$ in the outermost layer is in agreement with the (slightly off-axis) standard and choked jet models in our study. 
These two jet models assume a powerful jet lasting for $t_\mathrm{jet}=20\,\mathrm{s}$ and realize a high mass fraction of the iron-peak elements, $X_1\simeq 0.1$ out to $v\simeq c$ (standard jet) and $v\simeq 0.2c$--$0.3c$ (choked jet). 
On the other hand, the weak jet model with $t_\mathrm{jet}=4\,\mathrm{s}$ also produces high-velocity ejecta, but fails to contaminate it with the nucleosynthesis products, resulting in a much smaller mass fraction of the iron-peak element, $X_1=10^{-3}$, at larger radii. 
Therefore, a long-lasting jet capable of penetrating the massive star better explains the spectral evolution of the two GRB-SNe 2016jca and 2017iuk with the early spectra available. 

\subsubsection{GRB-SNe engines}
From the quantitative comparisons between our simulations and the observationally constrained outer density and chemical structure of GRB-SNe in the previous section, we argued that our jet models agree with the requirements; (1) a shallow radial density profile ($\rho\propto v^{-6}$) as opposed to normal SN ejecta ($\rho \propto v^{-10}$) and (2) highly abundant iron-peak elements with the mass fraction as much as $\sim 0.01$. 
This finding would suggest that both of an almost instantaneous energy deposition at the center and a long-lasting jet injection are required for some GRB-SNe with heavy elements in their early-time spectra. 
The former is required for producing enough amount of $^{56}$Ni powering the late-time optical emission. 
The latter component plays a role in transporting heavy nuclei produced in inner regions to the outermost layer.  
The earlier high energy deposition rate cannot last for a long time simply because of the available total energy budget. 
A long-lasting jet is also required to explain the properties of the prompt gamma-ray emission and the subsequent afterglow from individual GRBs. 

These two energy injection episodes should be related to each other.
As one possibility, these two may be caused by a single engine. 
For example, \cite{2018ApJ...860...38B} assumed a time-dependent jet injection rate in their end-to-end simulations of GRB-SNe. 
A high jet injection rate is initially set and then it decreases with time.  
Their models successfully reproduce the amount of $^{56}$Ni and the late-time spectral properties of GRB-SNe. 
However, detailed early-time spectral modelings and the comparison with observed GRB-SN spectra have not been attempted yet. 
As an alternative scenario, separated mechanisms may produce these two energy injection episodes. 
In the collapsar scenario \citep{1999ApJ...524..262M}, while the neutrino emission supposedly powers the jet, the associated disk wind may play a role in blowing out the stellar envelope and synthesizing heavy elements \citep{2005ApJ...629..341K,2008MNRAS.388.1729K,2018ApJ...854...43H}. 
More recently, \cite{2021arXiv210204467F} found a delayed onset of the neutrino-driven explosion in a collapsing rotating massive star in their simulations. 
In the proto-magnetar scenario, the prompt energy injection can be realized as a magneto-rotational explosion \citep[e.g.,][]{2003ApJ...584..954A,2006A&A...450.1107O,2009ApJ...691.1360T,2010ApJS..191..439K,2014ApJ...785L..29M,2015Natur.528..376M,2014MNRAS.445.3169O,2020ApJ...896..102K,2020MNRAS.492.4613O,2021MNRAS.500.4365A}, which is followed by a more gradual energy injection via the magnetic breaking of the rotating proto-magnetar. 

The jet-induced chemical mixing in GRB-SNe indicates that nuclei freshly synthesized in the explosions manifest themselves as absorption components in the early-time spectra of GRB-SNe. 
SNe from the collapsing very massive stars have been considered as a promising source of some nuclei, such as Zn \citep{2002ApJ...565..385U}, that are less likely supplied by normal CCSNe. 
Based on several post-process explosive nucleosynthesis calculations, it has been pointed out that magneto-hydrodynamic jet-driven explosions could be a possible $r$-process site \citep[e.g.,][]{2015ApJ...810..109N,2017ApJ...836L..21N,2018ApJ...864..171M,2021MNRAS.501.5733R}. 
Collapsar disks have also been considered as a possible $r$-process site \citep[e.g.,][]{2005ApJ...629..341K,2019Natur.569..241S,2020MNRAS.499.4097Z}. 
Although identifying specific elements with small mass fractions like rare iron-peak and $r$-process elements in an SN spectrum is practically difficult, they may collectively serve as an opacity source and hence make the spectrum redder. 
Therefore, spectral formation studies based on multi-dimensional simulation results \citep[e.g.,][]{2018ApJ...860...38B} and fair comparisons between observed and synthetic spectra are highly encouraged. 

\subsection{Spectral diversity of GRB-SNe}
The less efficient chemical mixing in the weak jet model results in relatively ``metal-clean’’ high-velocity ejecta. 
The early spectra of a GRB-SN with such high-velocity ejecta would be less affected by the UV photon blocking. 
\cite{2019MNRAS.487.5824A} compared the spectra of GRB-SNe 1998bw and 2016jca, and pointed out that those of SN 2016jca suffer from a more severe UV suppression than SN 1998bw, which showed bluer colors. 
This spectral diversity may be caused by the different degrees of metal mixing in the ejecta. 
In other words, SN 1998bw could be explained by a short-lived jet that failed to penetrate the star, while a successful jet is accompanied by the explosion responsible for SN 2016jca.  
It is widely known that GRB 980425 associated with SN 1998bw was extremely underluminous and the subsequent radio afterglow was also faint as a GRB \citep{1998Natur.395..663K,1998Natur.395..670G}. 
This is consistent with the failed jet scenario as has already been suggested by previous studies. 
A caveat is that the spectral difference may instead be attributed to the different progenitors; the light curve analysis suggests a less massive progenitor for SN 2016jca than that of SN 1998bw \citep{2019MNRAS.487.5824A}.  
Nevertheless, the spectral diversity caused by the different jet injection conditions would still be a possible and appealing explanation. 

As well as the jet injection conditions, the viewing angle dependence should also be important. 
The distribution of the iron-peak elements in Figures \ref{fig:final_fiducial} and \ref{fig:final_weak} show highly anisotropic spatial distributions. 
The materials traveling around the equatorial plane are less polluted by heavy metals. 
The region with the high mass fraction of the iron-peak elements $X_1$ is confined around the jet axis (metal-rich outflow). 
Therefore, when observed from an off-axis viewing angle, the line-of-sight velocity of metal-rich outflow can be smaller than that with an on-axis viewing angle. 
This difference would influence the appearance of the blue-shifted absorption component of heavy elements in early spectra of GRB-SNe.

\subsection{CSM-powered emission}
Finally, we consider the thermal emission from the high-velocity ejecta with a particular focus on the case of the interaction with the dense CSM, which potentially produces a bright transient. 

\subsubsection{Shock breakout}
The high-velocity ejecta colliding with the dense CSM as in the choked jet model produces a large amount of the internal energy via the shock dissipation, which would subsequently escape as thermal radiation. 
At the moment of the jet and cocoon emergence from the photosphere in the CSM, the release of high-energy photons from the shocked gas would be accompanied \citep{2013ApJ...764L..12S,2015ApJ...807..172N}. 
As seen in Figure \ref{fig:radial_csm}, the outgoing kinetic power of the outermost layer at the jet emergence ($t=10^3\,\mathrm{s}$) is as large as $\simeq 10^{50}\,\mathrm{erg}\,\mathrm{s}^{-1}$ along the jet axis $\theta=0^\circ$ and $\simeq 10^{48}\,\mathrm{erg}\,\mathrm{s}^{-1}$ even at $\theta=20^{\circ}$. 

We estimate the available radiation energy and the time scale of the shock breakout emission \citep[see][for a review]{2017hsn..book..967W}. 
We suppose that the forward shock driven by the jet emerges from the outer edge of the CSM. 
Assuming the equipartition between the kinetic and internal energies immediately after the shock passage, we expect the intrinsic (isotropic) luminosity of the shock breakout emission is roughly given by the outgoing kinetic power at the breakout radius $r=R_\mathrm{bo}$;
\begin{equation}
    L_\mathrm{bo}\simeq 4\pi \epsilon R_\mathrm{bo}^2\rho_\mathrm{bo}c^2\Gamma_\mathrm{bo}(\Gamma_\mathrm{bo}-1)v_\mathrm{bo},
\end{equation}
where $v_\mathrm{bo}$ and $\Gamma_\mathrm{bo}$ are the shock velocity and the corresponding Lorentz factor at the breakout and $\rho_\mathrm{bo}$ is the CSM density at the breakout radius,  $\rho_\mathrm{bo}=\rho_\mathrm{csm}(R_\mathrm{bo})$. 
The numerical factor $\epsilon$ represents the conversion efficiency of the shock kinetic energy into the breakout emission energy and is assumed to be of the order of unity. 
At the moment of the shock breakout, the optical depth of the thin layer ahead of the shock is assumed to be equal to the speed of light divided by the shock velocity; $\tau=c/v_\mathrm{bo}$. 
Since the optical depth of the layer with a width $L$ is given by $\tau=\kappa \rho_\mathrm{bo}L$, the breakout emission time scale $t_\mathrm{bo}$, which is roughly given by the shock crossing time of the layer, is estimated to be
\begin{equation}
    t_\mathrm{bo}=\frac{L}{v_\mathrm{bo}}
    =\frac{\tau}{\rho_\mathrm{bo}\kappa v_\mathrm{bo}}
    =\frac{c}{\rho_\mathrm{bo}\kappa v_\mathrm{bo}^2}.
\end{equation}
Therefore, the available (isotropic) radiation energy $E_\mathrm{bo}$ of the shock breakout emission is given by
\begin{equation}
    E_\mathrm{bo}\simeq L_\mathrm{bo}t_\mathrm{bo}=
    \frac{4\pi \epsilon cR_\mathrm{bo}^2v_\mathrm{bo}\Gamma_\mathrm{bo}^3}
    {\kappa(\Gamma_\mathrm{bo}+1)}
\end{equation}
With $R_\mathrm{bo}=3\times 10^{13}\ \mathrm{cm}$, $v_\mathrm{bo}=0.9c$, and $\kappa=0.2\ \mathrm{cm}^2\mathrm{g}^{-1}$, the radiation energy yields  $E_\mathrm{bo}\simeq 1.7\times 10^{50}\epsilon\ \mathrm{erg}$. 

The breakout photons would be predominantly emitted around the jet axis (the inclination angle within $\theta=20$--$30$ degrees), because of the elongated shape of the shocked gas (see Figure \ref{fig:evol_csm}). 
After the shock emergence around the jet axis, the laterally expanding gas quickly covers the equatorial part of the CSM in several $R_\mathrm{bo}/v_\mathrm{bo}\simeq 1000\ \mathrm{s}$, and therefore the breakout emission from shocks along large inclination angles is not expected. 
An observer sees the breakout photons from the emitting region with the length scale of $R_\mathrm{bo}/\Gamma_\mathrm{bo}^2$ due to the relativistic beaming. 
For $v_\mathrm{bo}=0.9c$ ($\Gamma_\mathrm{bo}\simeq 2.3$), the light crossing time across the emitting region is $R_\mathrm{bo}/(\Gamma_\mathrm{bo}^2c)\simeq 200\ \mathrm{s}$. 
The observed breakout emission would be smeared out within this light crossing time, $t_\mathrm{obs}\simeq 200\ \mathrm{s}$. 
Therefore, the observed luminosity of the breakout emission is given by
\begin{equation}
    L_\mathrm{bo,obs}\simeq\frac{E_\mathrm{bo}}{t_\mathrm{obs}}\simeq 10^{48}\epsilon\ \mathrm{erg}\ \mathrm{s}^{-1}.
\end{equation}

The energy spectrum of the breakout radiation is difficult to estimate. 
As seen in Figure \ref{fig:evol_csm} (2nd and 3rd rows), the post-shock pressure of the jet is of the order of $P=10^{10}\,\mathrm{erg}\,\mathrm{cm}^{-3}$ immediately before and after the breakout. 
The complete thermal equilibrium between matter and radiation leads to the radiation temperature of $T_\mathrm{ph}=(3P/a_\mathrm{r})^{1/4}\simeq 10^6\,\mathrm{K}\simeq 0.1\,\mathrm{keV}$. 
In the shock breakout at relativistic regime, however, the complete coupling between radiation and matter is not expected {in the immediate downstream of the shock \citep[see][for a review]{2017hsn..book..967W}. 
Therefore, the photon mean energy can be elevated to X-ray or maybe gamma-ray energy regime. 
If it is the case, the jet or cocoon emergence from the CSM potentially accounts for some X-ray flashes or low-luminosity GRBs, such as GRB 060218/SN 2006aj \citep{2006Natur.442.1008C,2006Natur.442.1018M,2006ApJ...645L..21M,2006Natur.442.1014S} and GRB 100316D/SN 2010bh \citep{2011MNRAS.411.2792S,2011ApJ...740...41C,2012ApJ...753...67B,2012A&A...539A..76O}. 
We also note that thermal soft X-ray emission with the photon temperature of $\sim 0.1\,\mathrm{keV}$ is identified in some GRBs \citep{2012MNRAS.427.2950S,2012MNRAS.427.2965S,2013ApJ...771...15F,2018MNRAS.474.2401V,2021MNRAS.501.4974V}, which may be attributed to the cocoon emission \citep{2013ApJ...764L..12S}.

\subsubsection{Post-shock breakout cooling emission}

After the shock breakout, thermal emission from the high-velocity ejecta continues but is dominated by lower-energy photons, as the ejecta expands and becomes transparent to thermal photons. 
This is an analog of the post-shock breakout cooling emission in canonical SNe. 
We consider a potential link between the thermal emission and an emerging population of peculiar optical transients detected by recent transient surveys \citep[e.g.,][]{2014ApJ...794...23D,2016ApJ...819...35A,2018MNRAS.481..894P,2020ApJ...894...27T,2021arXiv210508811H}. 

The population of luminous and fast optical transients includes a few events with late-time spectra consistent with SNe Ic-BL. 
The luminous and fast transient iPTF16asu \citep{2017ApJ...851..107W,2019MNRAS.489.1110W} exhibits unprecedentedly rapid evolution in the population of SNe Ic-BL. 
The early spectra were initially characterized by a blue continuum and then transformed into those of SNe Ic-BL. 
The fast-evolving light curve of iPTF16asu showed the rise time of $\sim 3\,\mathrm{days}$, the peak bolometric luminosity of $\sim 3\times 10^{43}\,\mathrm{erg}\,\mathrm{s}^{-1}$, and the total radiated energy of $\sim 4\times 10^{49}\,\mathrm{erg}$. 
SN 2018gep is another SN Ic-BL with rapid and bright optical emission \citep{2019ApJ...887..169H,2020arXiv200804321P}. 
The rise time of the optical emission was comparable to that of iPTF16asu, $\sim 3\,\mathrm{days}$, but it reached a much brighter bolometric peak luminosity of $>3\times 10^{44}\,\mathrm{erg}\,\mathrm{s}^{-1}$. 

One of the promising scenarios explaining the early luminous and blue emission from these rapidly evolving SNe Ic-BL is the shock cooling emission \citep[e.g.,][]{2010ApJ...725..904N,2014ApJ...788..193N,2015ApJ...808L..51P}, i.e., the thermal emission from the shocked extended envelope. 
\cite{2017ApJ...851..107W} suggest the existence of $\sim 0.5\,M_\odot$ of an extended material with the outer radius of $\sim 2\times 10^{12}\,\mathrm{cm}$ for iPTF16asu based on its rise time and peak luminosity. 
A smaller and even more extended material with a mass of $0.02\,M_\odot$ and an outer radius of $3\times 10^{14}\,\mathrm{cm}$ is suggested for SN 2016gep \citep{2019ApJ...887..169H}. 
Our choked jet model in fact realizes such an energy deposition into the extended material around the star. 
A part of the jet energy of $E_\mathrm{jet}\sim 10^{52}\,\mathrm{erg}$ is deposited into a $0.1\,M_\odot$ CSM with an outer radius of $3\times 10^{13}\,\mathrm{cm}$. 
As a result, a high-velocity ejecta with $M(\Gamma\beta>0.1)=0.06\,M_\odot$, $E_\mathrm{kin}(\Gamma\beta>0.1)=3.6\times 10^{51}\,\mathrm{erg}$, and $E_\mathrm{int}(\Gamma\beta>0.1)=8\times 10^{50}\,\mathrm{erg}$ is created at $t=10^{4}\,\mathrm{s}$ (see Figure \ref{fig:spec}).

In the following, we briefly consider the expected thermal radiation from the high-velocity ejecta. 
We assume a freely expanding spherical ejecta with the mass $M_\mathrm{ej}$. 
A power-law density profile with the minimum velocity of $v_\mathrm{br}$ and the power-law index $n$ ($>3$) is employed,
\begin{equation}
    \rho_\mathrm{ej}(t,r)=\frac{(n-3)M_\mathrm{ej}}{4\pi(v_\mathrm{br}t)^3}
    \left(\frac{v}{v_\mathrm{br}}\right)^{-n}.
\end{equation}
The optical depth of the ejecta is given by
\begin{equation}
    \tau=\int\kappa \rho_\mathrm{ej}dr=\frac{n-3}{4\pi(n-1)}
    \frac{\kappa M_\mathrm{ej}}{(v_\mathrm{br}t)^2},
\end{equation}
for a constant opacity $\kappa$. 
For the power-law exponent of $n=5$, the optical depth decreases with time as follows,
\begin{equation}
    \tau=1.6
    \left(\frac{\kappa}{0.2\,\mathrm{cm}^2\mathrm{g}^{-1}}\right)
    \left(\frac{M_\mathrm{ej}}{0.06\,M_\odot}\right)
    \left(\frac{v_\mathrm{br}}{0.1c}\right)^{-2}
    \left(\frac{t}{3\,\mathrm{days}}\right)^{-2}.
\end{equation}
Therefore, a $0.06\,M_\odot$ ejecta with $v_\mathrm{br}=0.1c$, which corresponds to the high-velocity ejecta in the choked jet model, becomes transparent within $3$--$4$ days. 
Assuming adiabatic cooling with $\gamma=4/3$, the internal energy of the ejecta decreases by a factor of $\sim 30$ from $t=10^{4}$ to $t=3$--$4\,\mathrm{days}$, resulting in $E_\mathrm{int}(\Gamma\beta>0.1)\sim 2\times 10^{49}\,\mathrm{erg}$. 
The internal energy at the epoch is still radiation-dominated and therefore most of the internal energy would be radiated away in $3$--$4\,$days after the explosion, resulting in the bolometric luminosity of $\simeq 6\times 10^{43}\,\mathrm{erg}\,\mathrm{s}^{-1}$. 
The time scale, the radiation energy, and the luminosity are comparable to those of iPTF 16asu \citep{2017ApJ...851..107W}. 
For SN 2018gep, the potential radiation energy of $2\times 10^{49}\,\mathrm{erg}$ at $3$--$4\,$days is smaller than the required radiation energy of $\sim 10^{50}\,\mathrm{erg}$, thereby implying an even more extended material with an outer radius of $>10^{14}\,\mathrm{cm}$ to overcome the severe adiabatic loss. 
Although we have only examined a single model,  extended materials with a wide variety of masses and outer radii appear to be required for explaining the population of rapid and luminous SNe Ic-BL. 
We leave such a parameter exploration as one of the future works.

\section{Summary}\label{sec:summary}

In this work, we have performed 3D simulations of phenomenological GRB jet propagation in a massive progenitor. 
We compare the three representative jet models, standard, weak, and choked jets to clarify their roles in forming the high-velocity ejecta. 
We found that the cocoon components in the three models show profound differences in the chemical structure, which can be probed by early-time spectroscopic observations of GRB-SNe, which has become available only recently. 
Early-time observations of GRB-SNe 2016jca and 2017iuk, and the spectral synthesis studies revealed a shallow radial density profile and abundant iron-peak elements in the sub-relativistic outer layers of the ejecta \citep{2019MNRAS.487.5824A,2019Natur.565..324I}. 
These features require depositing energy into the outer layers more efficiently than normal SNe and transporting explosive nucleosynthesis products into the outer layers. 
In this work, we have shown that:
\begin{itemize}
    \item A powerful jet penetrating a massive star, as is usually assumed for a GRB, can create high-velocity ejecta whose density and velocity structures are consistent with those suggested by observations; 
    \item Such a jet can also enrich the outer layers of the ejecta with explosive nucleosynthesis products, which makes the chemical abundance distinguished from that of the CO layer of a massive star; and
    \item Different jet configurations (standard, failed, and chocked) lead to a variety of early spectroscopic signals, which can be proved by future observations of GRB-SNe. 
\end{itemize}
These findings reinforce the idea that GRB 171205A/SN 2017iuk, which was accompanied by under-luminous gamma-ray emission, likely had a powerful relativistic jet though it was off-axis. 
More quantitative comparisons should be made by calculating synthetic spectra based on the simulation results. 
We also suggest that the thermal emission from the cocoon-CSM interaction can explain the population of rapidly evolving SNe Ic-BL, such as iPTF16asu and SN 2018gep. 
Our findings demonstrate the potential of early-time spectroscopy of GRB-SNe and fast transients in pinpointing the jet activities and therefore warrant further investigations.

\acknowledgements
A.S. acknowledges support by Japan Society for the Promotion of Science (JSPS) KAKENHI Grant Number JP19K14770. 
K.M. is supported by JSPS KEKENHI Grant Number JP20H00174, JP20H04737, and JP18H05223.

\software{Matplotlib (v3.2.1; \citealt{2007CSE.....9...90H})
}



\bibliography{refs}{}
\bibliographystyle{aasjournal}



\end{document}